\documentclass[sigconf,10pt]{acmart}

\usepackage[switch]{lineno}  
\usepackage[table]{xcolor}
\usepackage{colortbl}

\usepackage[normalem]{ulem}
\usepackage{array}
\usepackage{subcaption}
\usepackage{makecell}

\usepackage[linesnumbered,lined,ruled]{algorithm2e}

\SetCommentSty{mycmtsty}

\usepackage{blindtext}
\usepackage{booktabs}
\usepackage{caption}
\usepackage{multirow}
\usepackage{graphicx}
\usepackage{float}
\usepackage{adjustbox}
\usepackage{enumitem}
\usepackage{algpseudocode}
\usepackage{pifont}
\usepackage{diagbox}
\usepackage{amsmath}
\DeclareMathOperator*{\argmin}{arg\,min}

\usepackage{placeins}   
\usepackage{balance} 
\usepackage{enumitem}
\usepackage{hyperref}

\newcommand{\codename}{\textsf{SparseLoom}\xspace}
\usepackage{tikz}
\newcolumntype{P}[1]{>{\centering\arraybackslash}p{#1}}
\newcommand{\cmark}{\ding{51}}%
\newcommand{\xmark}{\ding{55}}%

\renewcommand\footnotetextcopyrightpermission[1]{}             

\newcommand{\out}[1] {}

\newcounter{codeLineCntr}

\reversemarginpar
\newif\ifnotes
\notestrue

\newcommand{\punt}[1]{}

\renewcommand{\eqref}[1]{Equation~(\ref{#1})}

\newcommand{\proc}[1]{\ifmmode\mbox{\textsc{#1}}\else\textsc{#1}\fi}

  \newcommand{\func}[1]{\ifmmode\mathrm{#1}\else\textrm{#1}fi} %

\newcounter{remark}[section]

\usepackage{xcolor}
\usepackage{tcolorbox}

\tcbset{
    sharp corners,
    colback = white,
    before skip = 0.3cm,    
    after skip = 0.25cm      
}                           

\newtcolorbox{boxH}{
    colback = gray!8, 
    colframe = black!60,
    boxrule = 0.5pt, 
    leftrule = 3pt,
    left=2pt,
    right=2pt,
    top=2pt,
    bottom=2pt
}

\begin{document}


\title[\codename]{Multi-DNN Inference of Sparse Models on Edge SoCs}


\author{Jiawei Luo, Di Wu, Simon Dobson, Blesson Varghese}

\affiliation{
  School of Computer Science, University of St Andrews, Scotland UK
}
\email{{jl425, dw217, simon.dobson, bv6}@st-andrews.ac.uk}



\begin{abstract}
Modern edge applications increasingly require multi-DNN inference systems to execute tasks on heterogeneous processors, gaining performance from both concurrent execution and from matching each model to the most suited accelerator.
However, existing systems support only a single model (or a few sparse variants) per task, which impedes the efficiency of this matching and results in high Service Level Objective violation rates.
We introduce \emph{model stitching} for multi-DNN inference systems, which creates model variants by recombining subgraphs from sparse models without re-training.
We present a demonstrator system, \codename, that shows model stitching can be deployed to SoCs.
We show experimentally that \codename reduces SLO violation rates by up to 74\%, improves throughput by up to 2.31$\times$, and lowers memory overhead by an average of 28\% compared to state-of-the-art multi-DNN inference systems.
\end{abstract}


\maketitle


\section{Introduction}
\label{sec:introduction}

\begin{figure}[t]
    \centering    \includegraphics[width=1.0\linewidth]{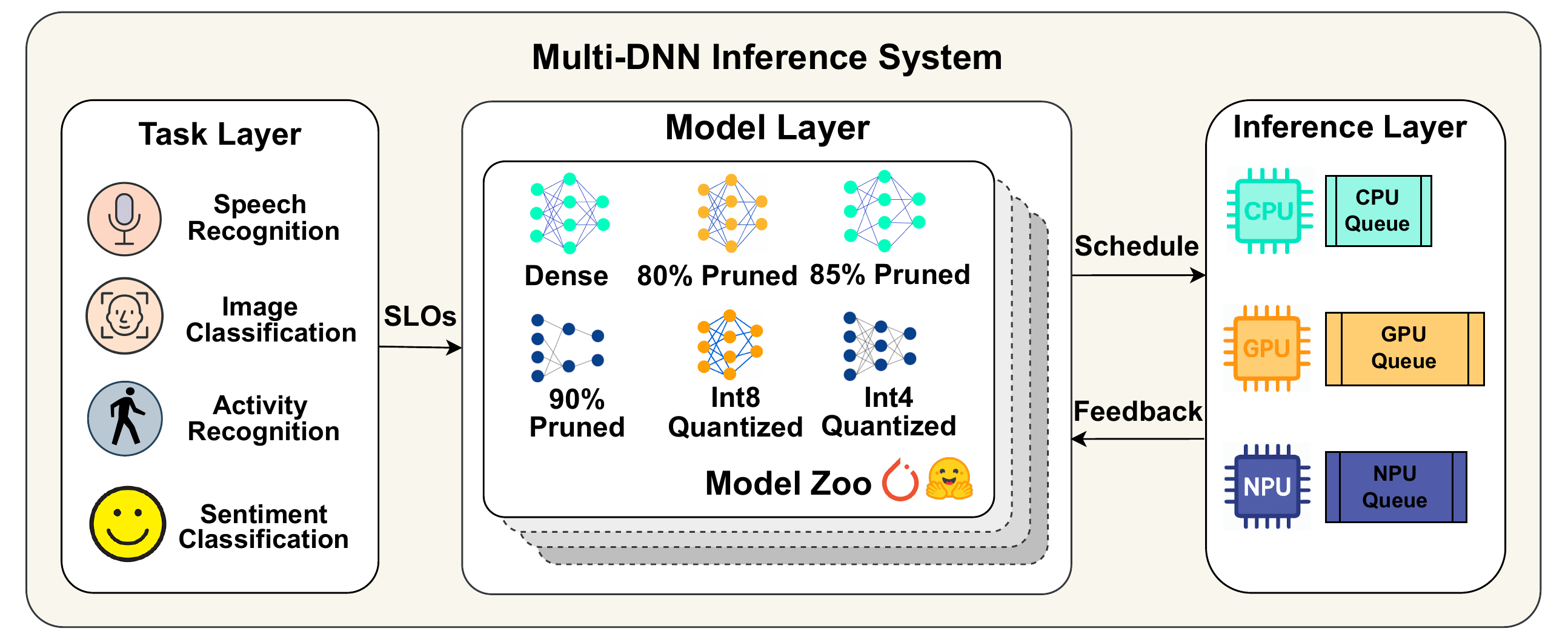}
    \caption{An AR use case of multi-DNN inference on a heterogeneous edge SoC. The application executes four tasks in parallel on the CPU, GPU, and NPU. Each task is provisioned with a sparse model zoo to meet varying SLO requirements.}
    \label{fig:architecture}
\end{figure}

Modern edge applications often need to run diverse tasks in parallel~\cite{iyer2023automated, han2022microsecond,jeong2022band}. For example, an augmented reality (AR) application enhances user experience by executing multiple concurrent tasks (as illustrated in the task layer of Figure~\ref{fig:architecture}), including speech recognition~\cite{mirzaei2014audio}, image classification~\cite{thilahar2019fuzzy,satyanarayanan2019augmenting}, activity recognition~\cite{ouyang2022cosmo}, and sentiment classification~\cite{mondal2024analysing}.
Each of these deep learning tasks places different constraints on the underlying hardware.

Furthermore, there may be multiple Service Level Objectives (SLOs) constraining the system - for example, the latency/accuracy priorities may change on an ongoing basis~\cite{han2024pantheon,ling2021rt,crankshaw2020inferline,ahmad2024loki,ahmad2024proteus}.
Each task can therefore be provisioned from multiple sparse variants (as illustrated in the model layer of Figure~\ref{fig:architecture}). These sparse variants are derived from a common base model \textit{via} pruning~\cite{cheng2024survey,yang2023global} or quantization~\cite{gholami2022survey,yuan2022ptq4vit}. Collectively, these variants form a \textit{sparse model zoo} that supports diverse SLO requirements. For example, image classification in AR may require low latency for real-time AR labeling but may tolerate higher latency for background scene analysis. Deploying different variants matching these requirements can improve overall system utilization and application performance.

At the inference layer, these models are typically deployed on edge System-on-Chips (SoCs) comprising Central Processing Units (CPUs), Graphics Processing Units (GPUs), and Neural Processing Units (NPUs). To fully exploit the computational capabilities of these processors, models are often partitioned into subgraphs and distributed to the heterogeneous processors for parallel execution, thereby improving overall inference throughput~\cite{jeong2022band,karatzas2023omniboost,hetero2pipe2025}.


The multi-DNN inference system requires selecting and scheduling suitable sparse variants on the available devices, and re-optimizing selection and placement based on SLO feedback. \textbf{Ideally such a system should minimize SLO violation rates and maximize throughput when serving multiple concurrent tasks with diverse SLOs.}




\begin{figure}[t]
    \centering
    \includegraphics[width=1.0\linewidth]{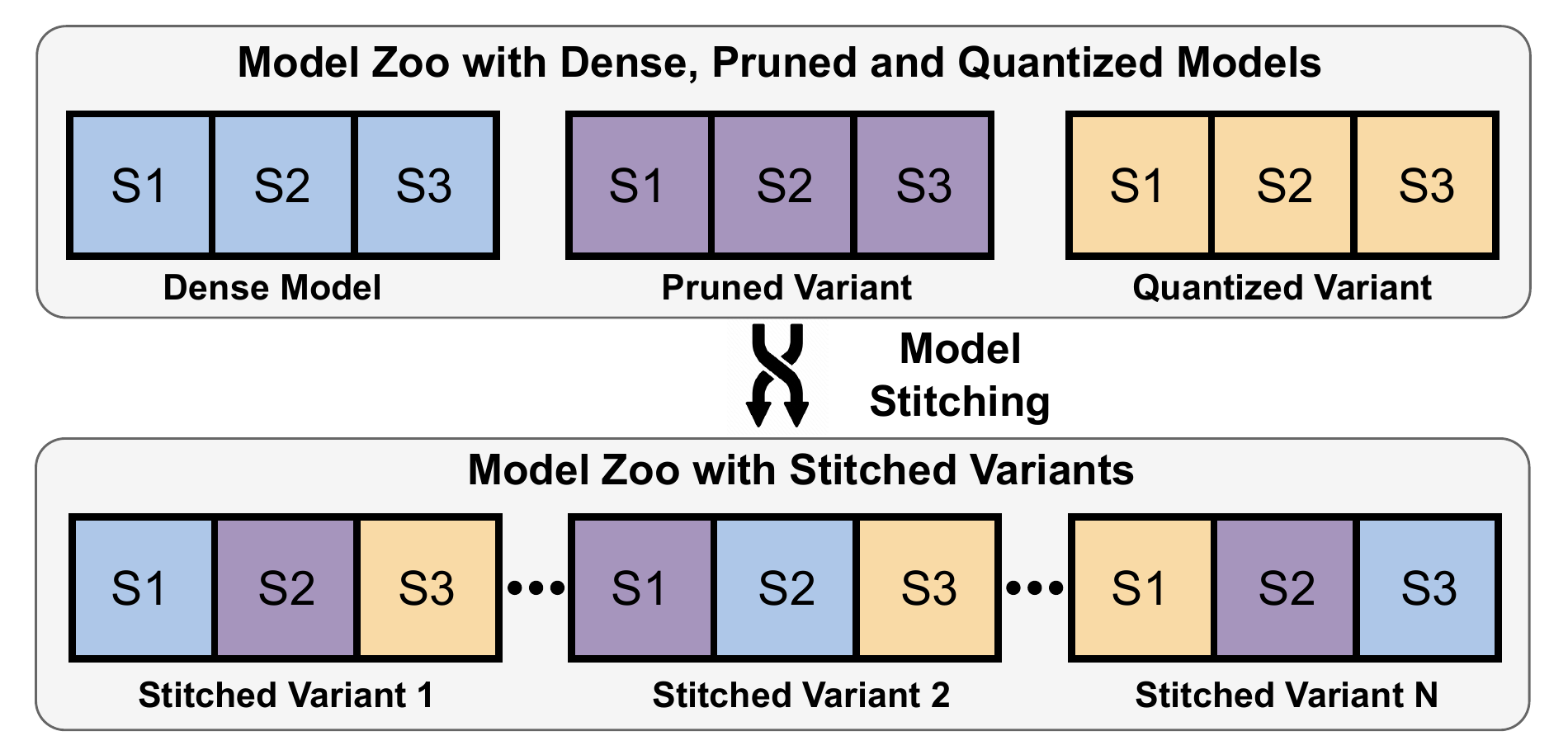}
    \caption{
    Model stitching: given three sparse models (dense, pruned, quantized) each split into subgraphs S1–S3, Stitched Variant 1 combines S1 from the Dense Model (blue), S2 from the Pruned Variant (purple), and S3 from the Quantized Variant (orange).}
    \label{fig:Model_Stitching_Method}
\end{figure}


Existing multi-DNN inference systems can be categorized based on whether they support sparse model selection. The majority of existing works assume a single base model per task and cannot adaptively select variants from a sparse model zoo~\cite{wang2019high,han2024pantheon,han2022microsecond}. These systems fail to leverage diverse accuracy-latency trade-offs in meeting varying SLO constraints. Recent works incorporate a model selector based on techniques such as reinforcement learning~\cite{taufique2024tango,sen2024elastically,fang2018nestdnn} to support adaptive variant selection. However, our evaluation shows that only selecting from the \textit{original} sparse variants often leads to high SLO violation rates due to the limited number of variants available (see Section~\ref{subsec:motivation}).

To reduce SLO violation rates in multi-DNN inference we introduce \textbf{model stitching}. 
Prior approaches apply stitching by analyzing individual layers~\cite{bansal2021revisiting} and creating variants at the expense of re-training~\cite{pan2023stitchable}.
In contrast, the proposed model stitching approach constructs stitched variants by combining subgraphs (\textit{i.e.}, consecutive layer blocks) from different sparse models \textit{without} re-training. 
These heterogeneous variants with diverse sparsity levels and patterns offer more options to satisfy task SLOs. 
As a result, \textbf{model stitching significantly reduces SLO violation rates without re-training} (see Section~\ref{subsec:motivation}).

Integrating model stitching in multi-DNN inference systems poses three challenges: (1) large profiling cost from the exponential number of possible stitched variants, (2) throughput degradation from suboptimal processor placement, and (3) substantial memory overheads from pre-loading stitched variants. 
We propose \codename, a holistic multi-DNN inference system for edge SoCs that makes model stitching deployable with lower profiling cost, higher throughput, and lower memory overhead.
A \textit{Performance Profiler} module employs an accuracy and latency estimator to predict the performance of stitched variants, thereby reducing the profiling cost.
A \textit{Sparsity-Aware Optimizer} module then selects the optimal processor placement order and determines stitched variants.
Finally, a \textit{Hot-Subgraph Preloader} module pre-loads subgraphs with high ``hotness'' scores for a given global memory budget, thereby improving memory utilization. Extensive evaluation of multiple target systems, models, and workloads shows that \codename reduces SLO violation rates by up to 74\%, improves throughput by up to 2.31$\times$, and lowers memory overhead by an average of 28\% compared to state-of-the-art multi-DNN inference systems. 



\section{Motivation and Challenges}
\label{sec:motivation}

In this section, we first empirically evaluate the performance of existing multi-DNN inference systems, with and without model stitching, to highlight its advantages in reducing SLO violations (Section~\ref{subsec:motivation}). 
We then identify three key challenges when incorporating model stitching into current systems (Section~\ref{subsec:challenges}), which motivate the design of \codename.

\textbf{Experimental setup:} 
The performance of current multi-DNN inference systems is evaluated on four representative models: ResNet101~\cite{he2016deep}, BERT-Base~\cite{devlin2019bert}, ViT-Small~\cite{dosovitskiy2020image}, and Wav2vec2~\cite{baevski2020wav2vec}, corresponding to image classification, sentiment classification, activity recognition, and speech recognition tasks, respectively. For each task, we construct a sparse model zoo consisting of ten sparsity variants, as detailed in Appendix~\ref{appendix:variants}.
The results shown are on an Intel Ultra 7 265K SoC, with an integrated CPU, GPU, and NPU~\footnote{The observations generalize across other hardware platforms (see Section~\ref{subsec:setup}).}.

\begin{figure}[tp]
    \centering
    \includegraphics[width=1\linewidth]{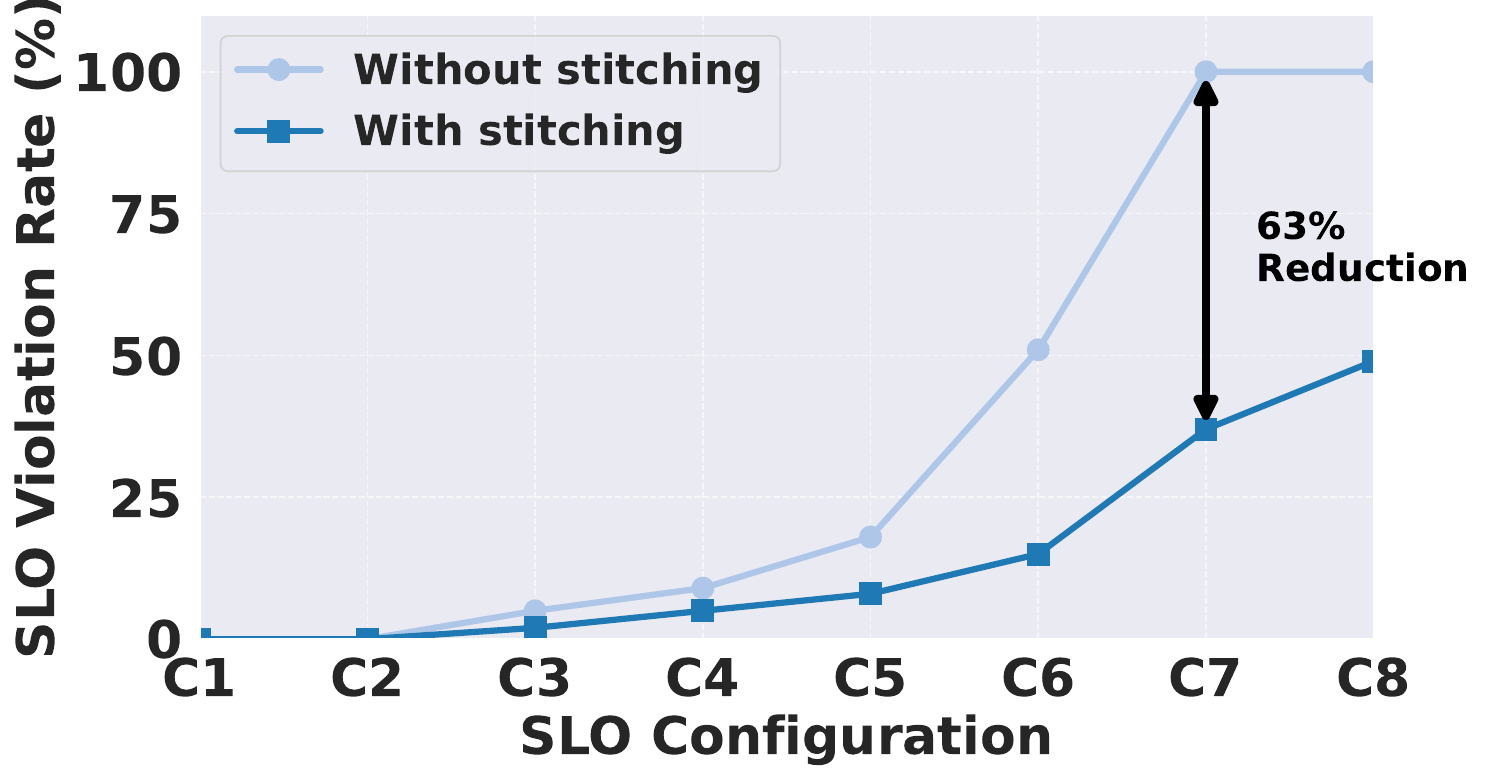}
    \caption{SLO violations with vs. without stitching. Stitching substantially reduces SLO violation rate. The x-axis represents different SLO configurations, where larger index (e.g., C8) corresponds to more challenging SLO with stricter accuracy and latency requirements.}
    \label{fig:8slo_violation_rate}
\end{figure}

\begin{figure}[tp]
    \centering
    \begin{subfigure}[t]{0.49\columnwidth}
        \centering
        \includegraphics[width=\linewidth]{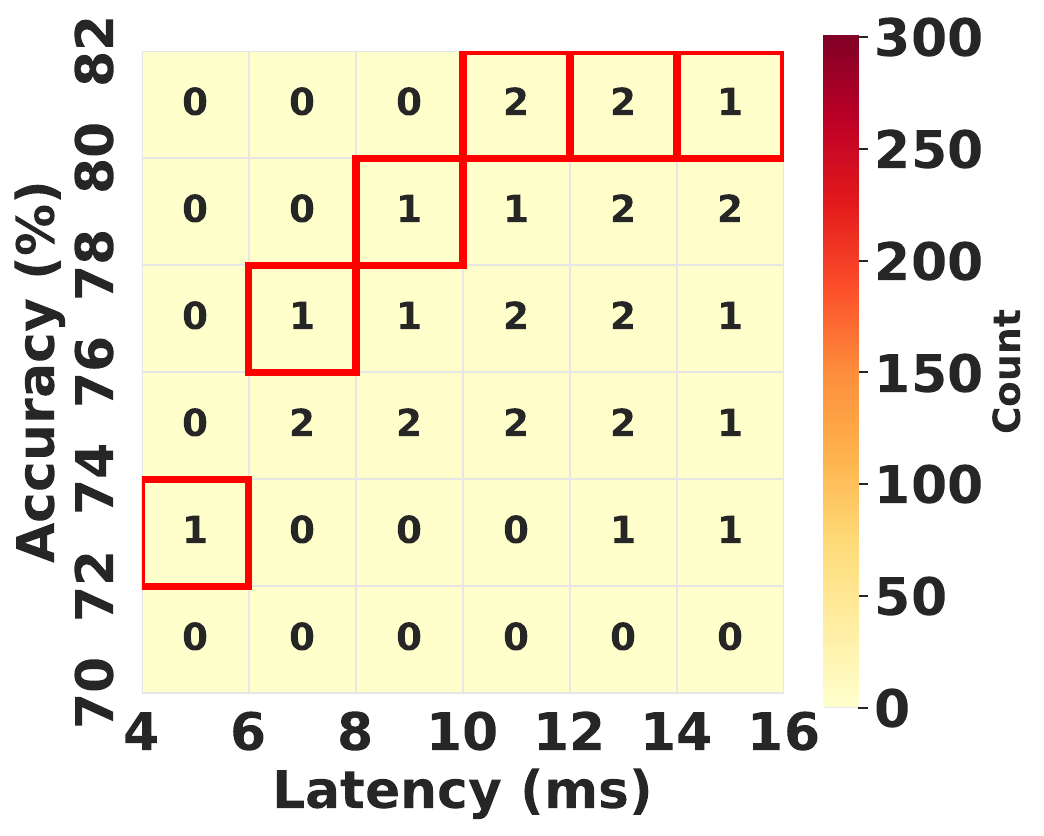}
        \caption{Without Stitching.}
        \label{fig:Sparse_Variants_Heatmap}
    \end{subfigure}
    \hfill
    \begin{subfigure}[t]{0.49\columnwidth}
        \centering
        \includegraphics[width=\linewidth]{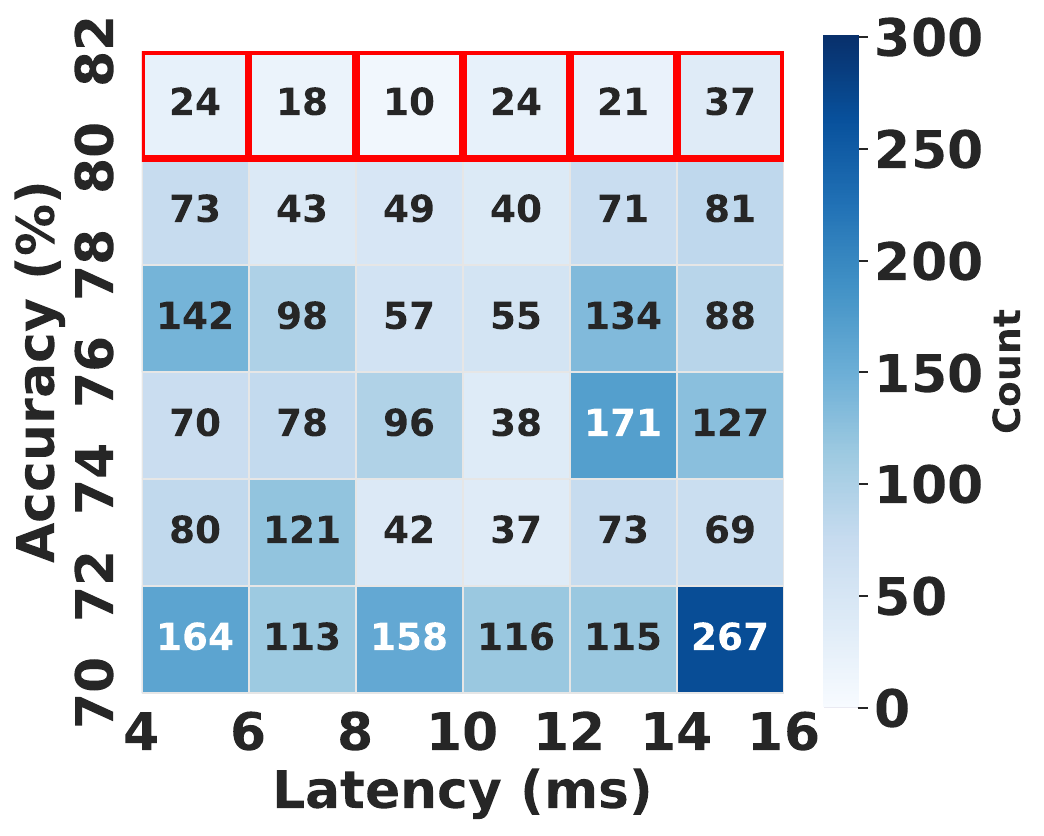}
        \caption{With Stitching.}
        \label{fig:Stitched_Variants_Heatmap}
    \end{subfigure}
    \caption{Histogram of ResNet101 variants in the accuracy–latency space; cell counts show density, and red-edged cells indicate the Pareto frontier. Model stitching expand the variant space and achieve a better accuracy–latency frontier than original sparse models.}
    \label{fig:variant_distribution}
\end{figure}

\subsection{Definition, Scope and Benefit of Model Stitching}
\label{subsec:motivation}
Existing multi-DNN inference systems typically select model variants directly from a sparse model zoo. These variants include the original dense model and its pruned and quantized versions. In contrast, \codename adopts model stitching to generate new variants.

\textbf{Definition:} \textbf{Model stitching is a training-free variant-generation technique that creates new model variants by combining subgraphs whose layers are aligned from sparse variants of the same base model.} Figure~\ref{fig:Model_Stitching_Method} illustrates model stitching, where subgraphs ($S1$–$S3$) from different consecutive layer blocks of models in the original model zoo are combined to form new stitched variants.

\textbf{Operational scope:}
(i) Sparse variants originate from the same base model through compression techniques (i.e., quantization and pruning),
and (ii) Sparse variants share an identical subgraph partitioning scheme to ensure that subgraphs can be recombined with aligned layer positions (i.e., layer-aligned subgraphs).

\textbf{Supported compression techniques and rationale:} Compression techniques of unstructured pruning (zero-masking), structured pruning (architecture-changing), and quantization, are effective for model stitching because they preserve the layer-wise output distribution across sparse variants. This is well studied in the model compression literature, as demonstrated by similarity-analysis theory (e.g., SVCCA~\cite{raghu2017svcca}). In addition, many compression methods explicitly enforce similar layer output distributions when reducing layer parameters, such as BRECQ~\cite{li2021brecq} and AMC~\cite{he2018amc}. Therefore, high accuracy can be maintained when stitching subgraphs from sparse variants compressed from the same base model.

\textbf{Benefits:} Model stitching enables the construction of new variants, thereby increasing the number of available variants for selection in a training-free manner.
\textit{However, there is limited understanding of whether model stitching can reduce the SLO violation rate compared to directly selecting from the original sparse models, as well as how it improves the accuracy–latency trade-offs.}

Figure~\ref{fig:8slo_violation_rate} shows the average SLO violation rates with and without model stitching across multiple tasks. Specifically, we benchmark the base model and all sparse variants to measure their accuracy and latency. We then extend these ranges by $\pm20\%$ for latency and $\pm2\%$ for accuracy, and uniformly sample eight SLO configurations (C1--C8) from the extended space. As the SLO becomes more stringent (from C1 to C8), the violation rate without model stitching increases, reaching 100\% under the strictest configuration (C8). However, under stricter SLOs, model stitching achieves up to a 63\% reduction in violation rate, with the largest improvement on the C7 configuration.

Figure~\ref{fig:variant_distribution} contrasts the distribution of original sparse variants (dense, pruned, and quantized) with stitched variants—both derived from ResNet101—in the accuracy–latency space. Model stitching not only increases the number of available variants but also achieves a better Pareto frontier (shown as red-edged cells), thereby offering improved accuracy–latency trade-offs for selection. Moreover, model stitching can even produce variants with higher accuracy or lower latency: 4\% of stitched variants exceed the highest accuracy of the original models, and 5\% achieve lower latency than the fastest model in the original sparse zoo.


\subsection{Challenges of Model Stitching}
\label{subsec:challenges}
In this section, we discuss the challenges that need to be surmounted in incorporating model stitching into multi-DNN inference systems.

\begin{boxH}
\textbf{Challenge 1:} \textit{
The large number of variants introduced by model stitching poses a significant scalability challenge to efficiently profiling their accuracy and latency.}
\end{boxH}

All variants (including models from original sparse model zoos and stitched variants) are profiled to record metrics, namely accuracy and latency. However, the profiling cost increases due to the large number of stitched variants.

Table~\ref{tab:profiling_cost} shows the total number of profiling runs required with and without model stitching. A profiling run refers to profiling a variant for both accuracy and latency. The processor placement order is the sequence in which subgraph variants are executed across processors. Each variant needs to be profiled under all processor placement orders.
Regarding the increase in profiling runs with the number of tasks $T$ and variants $V$: with model stitching, the number of profiling runs grows rapidly - linearly with the number of tasks $T$ and exponentially with the number of variants per task $V$. This highlights the scalability challenge posed by model stitching and motivates the need for more efficient profiling strategies.

After profiling all variants, the inference system selects a suitable variant for each task based on the SLO requirement. Specifically, the variant that can meet both the accuracy and latency constraints is chosen. To meet the latency constraint, the order in which the subgraphs must be placed on each processor needs to be considered as the execution time of each subgraph is highly dependent on the processor type.

\begin{boxH}
\textbf{Challenge 2:} \textit{
The processor placement order for executing subgraphs of stitched variants with diverse sparsity types is often latency-suboptimal, resulting in low overall throughput.
}
\end{boxH}


\begin{table}[tp]
\centering
\caption{Profiling complexity with and without model stitching. 
Notation: $T$: \#tasks; $V$: \#variants per task; $S$: \#subgraphs per variant; $P$: \#processors.}
\label{tab:profiling_cost}
\small
\renewcommand{\arraystretch}{1.2}
\begin{tabular}{P{3.5cm} P{1.9cm} P{1.9cm}}
\Xhline{2\arrayrulewidth}
\textbf{Method} & \textbf{Without Stitching} & \textbf{With Stitching} \\
\Xhline{2\arrayrulewidth}
Processor placement orders  & $P!$              & $P!$ \\
Total variants              & $T \cdot V$       & $T \cdot V^S$ \\
Accuracy profiling runs     & $T \cdot V$       & $T \cdot V^S$ \\
Latency profiling runs      & $T \cdot V \cdot P!$ & $T \cdot V^S \cdot P!$ \\
Total profiling runs        & $T \cdot V \cdot (P!+1)$ & $T \cdot V^S \cdot (P!+1)$ \\
\Xhline{2\arrayrulewidth}
\end{tabular}
\end{table}

\textbf{Placement order of processors:}
When stitched variants for multiple tasks are executed in parallel on heterogeneous processors, their subgraphs are typically assigned to processors in a fixed order~\cite{hetero2pipe2025,jeong2022band}. For example, given variants with three subgraphs, the system may consistently assign the first subgraph to the NPU, the second to the GPU, and the third to the CPU. This fixed placement strategy is adopted in existing multi-DNN inference systems to simplify placement and reduce the overhead of rescheduling during runtime~\cite{hetero2pipe2025}.

Table~\ref{tab:order_config_latency} presents the latency of executing a selection of stitched variants of ResNet101 with different processor placement orders. Given three processors (CPU, GPU, and NPU), there are six possible placement orders\footnote{Non-overlapping placement orders only to maximize parallelism.}. 
The results demonstrate that the optimal placement order varies across different sparse variants. More importantly, the widely adopted NPU–GPU–CPU (N-G-C) placement order~\cite{hetero2pipe2025,jeong2022band} consistently yields suboptimal latency, thereby degrading overall throughput (see Section~\ref{subsec:individual modules}). 

\textbf{Optimal processor placement order:}
Determining the optimal processor placement order is challenging because (i) it depends on the stitched variant chosen for each task, and (ii) different tasks may require different orders since different types of variants are selected.
Therefore, the processor placement order must be determined together with the final variant choices across all tasks. 
This motivates the need for automated and joint optimization of processor placement order and variant selection.

\begin{table}[tp]
\caption{Latency of stitched ResNet101 variants with different processor placement orders. The lowest latency is highlighted in blue. P: Pruned, Q: Quantized, D: Dense; C: CPU, G: GPU, N: NPU.}
\centering
\small
\setlength{\tabcolsep}{3pt}
\renewcommand{\arraystretch}{1.2}

\begin{tabular}{ccccccc}
\Xhline{2\arrayrulewidth}
\diagbox[width=5em]{\footnotesize\textbf{Order}}{\footnotesize\textbf{Variant}}
& \textbf{P-Q-P} & \textbf{P-P-Q} & \textbf{D-D-P} & \textbf{D-P-Q} & \textbf{Q-P-D} & \textbf{P-D-Q} \\
\Xhline{2\arrayrulewidth}
\textbf{N-G-C} & 12.05 & 16.91 & 14.77 & 17.73 & 18.25 & 16.99 \\
\textbf{C-G-N} & \cellcolor{cyan!20}11.01 & 13.40 & 14.45 & 15.56 & 20.27 & 13.48 \\
\textbf{G-C-N} & 13.20 & 13.69 & 13.51 & 12.14 & \cellcolor{cyan!20}12.17 & 15.54 \\
\textbf{G-N-C} & 12.98 & 14.22 & \cellcolor{cyan!20}13.49 & 14.57 & 13.63 & 16.51 \\
\textbf{N-C-G} & 15.72 & 11.93 & 17.39 & \cellcolor{cyan!20}12.01 & 13.79 & 15.73 \\
\textbf{C-N-G} & 13.72 & \cellcolor{cyan!20}10.77 & 15.40 & 12.88 & 18.21 & \cellcolor{cyan!20}12.51 \\
\Xhline{2\arrayrulewidth}
\textbf{Best Order} 
& \textbf{C-G-N} 
& \textbf{C-N-G} 
& \textbf{G-N-C} 
& \textbf{N-C-G} 
& \textbf{G-C-N} 
& \textbf{C-N-G} \\
\Xhline{2\arrayrulewidth}
\end{tabular}
\label{tab:order_config_latency}
\end{table}

\begin{boxH}
\textbf{Challenge 3:} \textit{
A substantial memory overhead is incurred when preloading all stitched variants to avoid runtime switching latency. 
}
\end{boxH}

\begin{figure}[tp]
    \centering
    \begin{subfigure}[t]{0.49\columnwidth}
        \centering
        \includegraphics[width=\linewidth]{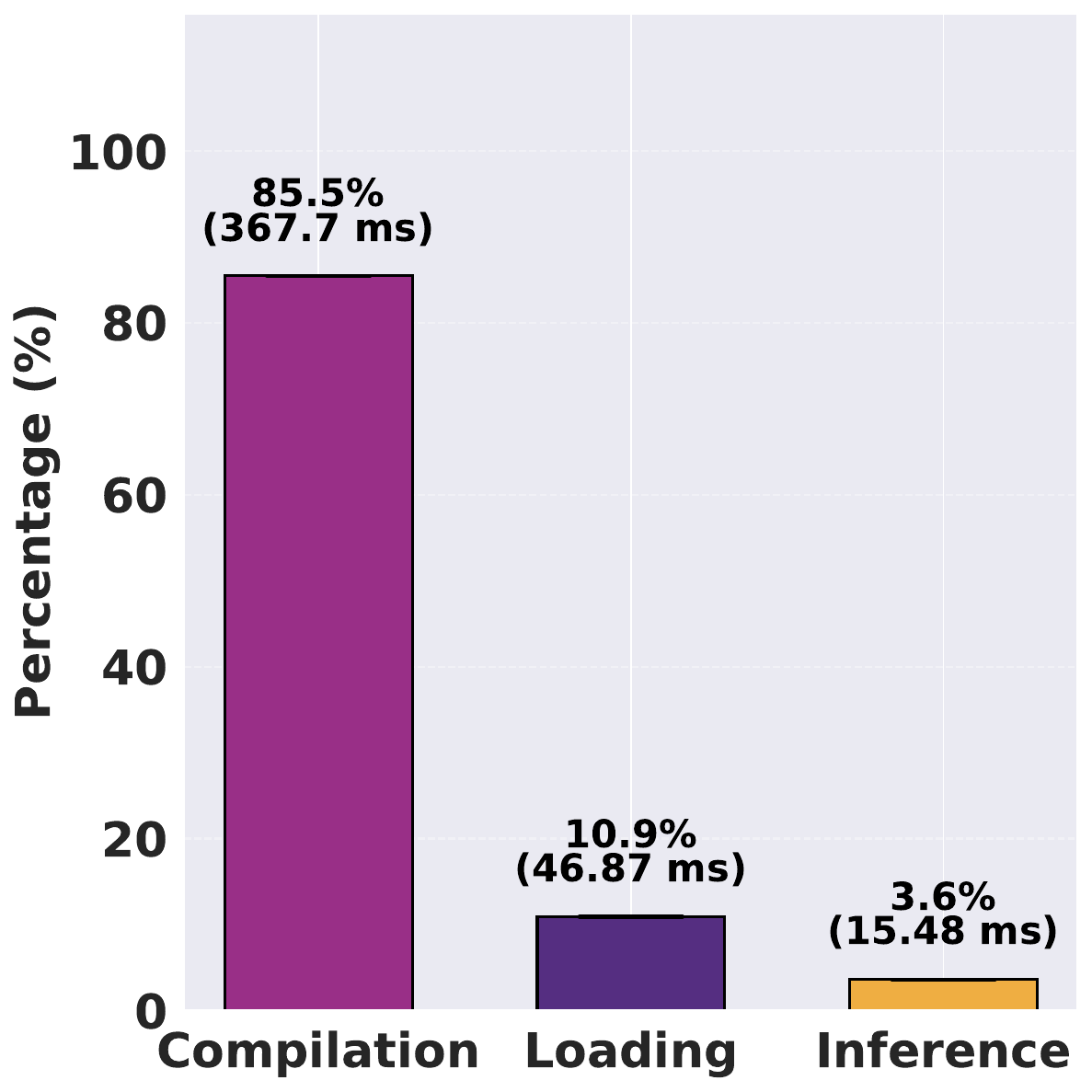}
        \caption{Latency Overhead.}
        \label{fig:lat_overhead}
    \end{subfigure}
    \hfill
    \begin{subfigure}[t]{0.49\columnwidth}
        \centering
        \includegraphics[width=\linewidth]{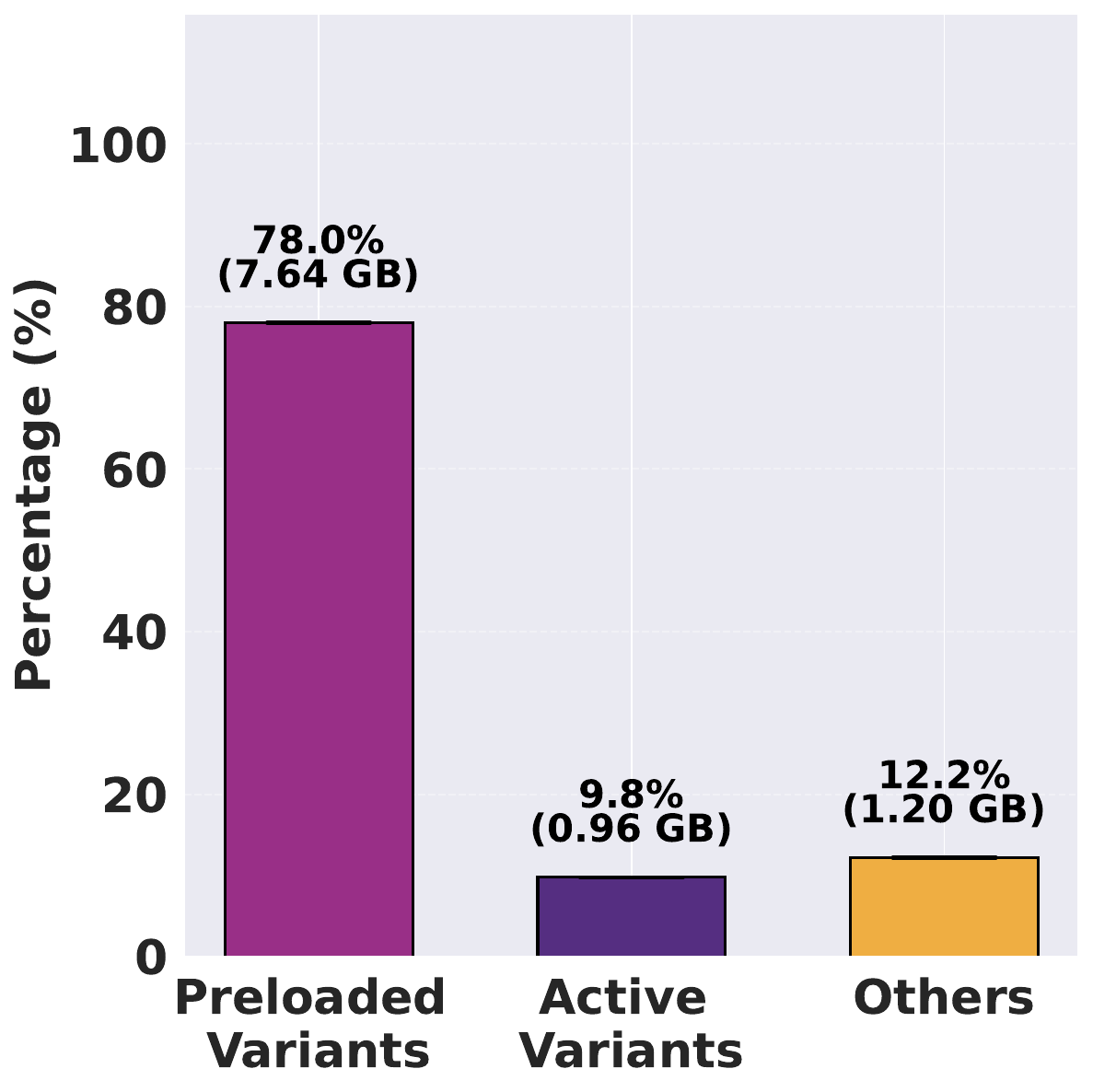}
        \caption{Memory Overhead.}
        \label{fig:mem_overhead}
    \end{subfigure}
    \caption{(a) Latency breakdown, including compilation, loading, and inference. (b) Memory breakdown, including active variants, preloaded variants, others.}
\end{figure}

\textbf{Latency overhead of switching variants:}
SLO violations are monitored during the execution of the scheduled tasks. If violation is detected, the system switches to a different variant in the model zoo to satisfy the SLO.
This includes the latency of compiling a new variant and loading it into the target processor’s memory.
Figure~\ref{fig:lat_overhead} compares the latency of compilation, loading, and inference when adding new variants. The loading overhead is substantial, accounting for up to 96.4\% of the total time. Specifically, compilation and loading take 23.7$\times$ and 3$\times$ longer than inference, respectively.

\textbf{Preloading the variants:}
To reduce switching overhead during runtime rescheduling, existing systems often pre-compile and preload all variants from disk into target memory~\cite{fang2018nestdnn,lee2020fast}. While this approach hides the switching latencies at runtime, it has a substantial memory requirement.
Figure~\ref{fig:mem_overhead} shows the runtime memory usage; substantial memory is required for preloading variants. This motivates the need for efficient preloading to switch variants.

In short, model stitching achieves lower SLO violations by offering a richer variant space, but it introduces non-trivial challenges that must be tackled to enable efficient deployment on multi-DNN inference systems.
\section{Design of \codename}
\label{sec:system}

\begin{figure}[tp]
    \centering    \includegraphics[width=1.0\linewidth]{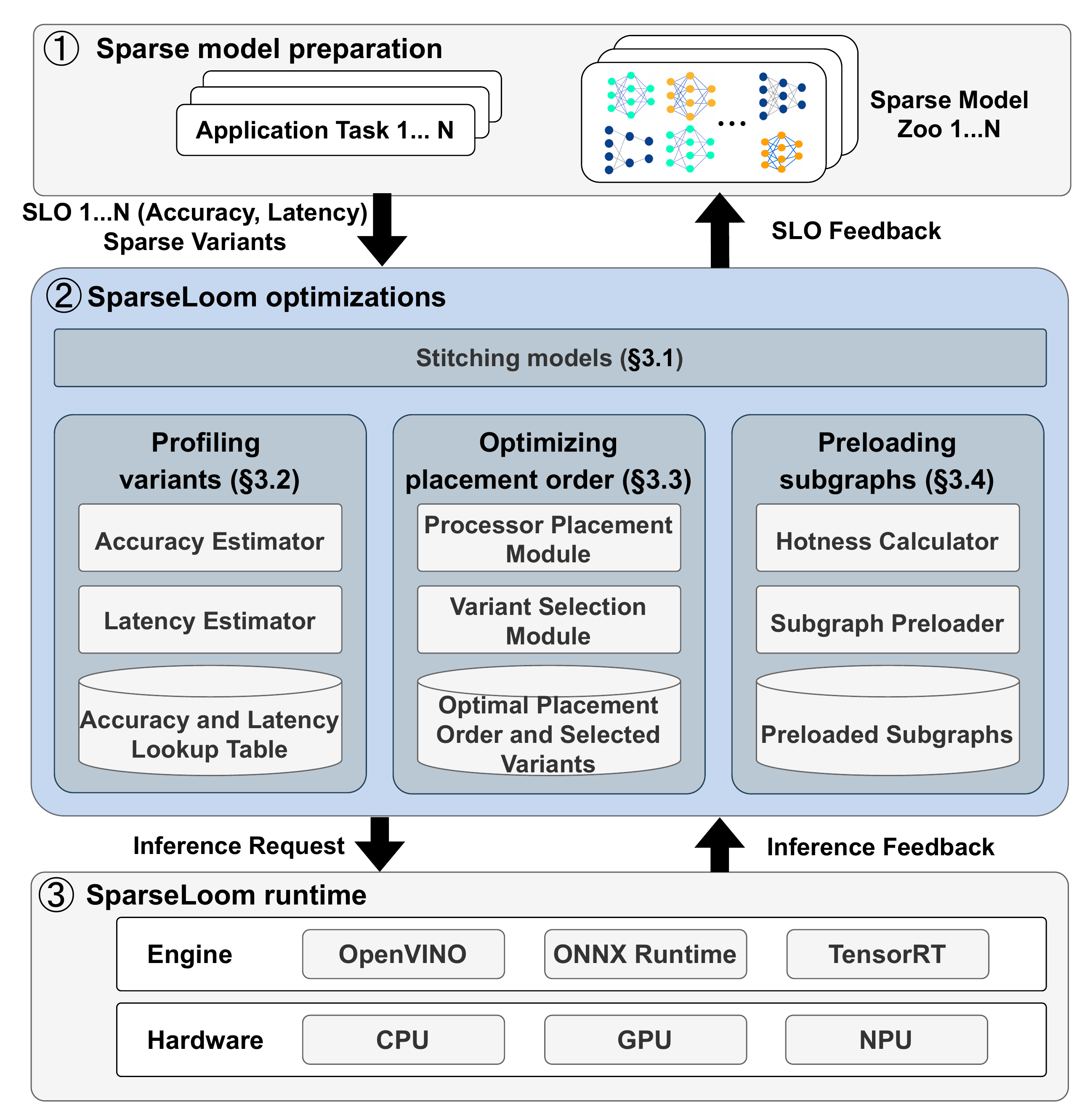}
    \caption{Three phases of multi-DNN inference in \codename.}
    \label{fig:SparseStitch_architecture}
\end{figure}

\codename introduces the core technique of \textit{model stitching}, supported by three novel modules that jointly address the challenges of high profiling cost, suboptimal processor placement, and memory-intensive subgraph preloading. 

Figure~\ref{fig:SparseStitch_architecture} illustrates the three phases of multi-DNN inference in \codename.
\textbf{\textcircled{1} Sparse model preparation:} Each task in the application is prepared with task-specific SLOs and a sparse model zoo generated by compressing the base model into different sparse variants.
\textbf{\textcircled{2} \codename optimizations:} \codename first stitches variants within each task’s sparse models (Section~\ref{subsec:stitching}). It then employs the accuracy and latency estimators to profile stitched variants, producing an accuracy–latency lookup table (Section~\ref{subsec:profiler}). 
Next, \codename determines an optimal processor placement order and selects the optimal stitched variant for each task (Section~\ref{subsec:optimizer}). 
To reduce the memory overhead of preloading, the subgraph preloader loads subgraphs based on a hotness metric, which captures both the frequency and uniqueness of subgraphs (Section~\ref{subsec:preloader}).
\textbf{\textcircled{3} \codename runtime:} 
After \codename determines an optimal processor placement order and selects the stitched variant for each task, it dispatches inference requests to the inference engines across multiple processors and collects feedback.

\subsection{Stitching subgraphs from sparse variants}
\label{subsec:stitching}
Each of the $T$ tasks is associated with $V$ sparse variants, where each variant $v^{t,i}$ is composed of $S$ subgraphs $\{s_j^{t,i}\}_{j=1}^S$. 
\codename constructs stitched variants in a training-free manner by combining subgraphs from these original variants. 
Each stitched variant $\tilde{v}^{t,k}$ also consists of $S$ subgraphs $\{\tilde{s}_j^{t,k}\}_{j=1}^S$, where each $\tilde{s}_j^{t,k}$ is inherited from a $s_j^{t,i}$. 
This results in a natural mapping between original and stitched variants, since the stitched variants directly reuse subgraphs from the originals. The mapping is formally defined as:
\begin{equation}
\tilde{s}_j^{t,k} = s_j^{t,M[j, i]}, \quad \text{where } k = M[j, i].
\label{eq:subgraphs_mapping}
\end{equation}
$M[j,i]$ denotes the mapping function, and specifies that in stitched variant $k$, the subgraph at position $j$ is taken from variant $i$ at the same position. The notation used in \codename is presented in Appendix~\ref{appendix:notation}.

\subsection{Estimating Accuracy and Latency of Stitched Variants}
\label{subsec:profiler}

\textbf{Accuracy and latency estimators: }
With the stitched variants prepared, \codename measures the accuracy $A(\tilde{v}^{t,k})$ and latency $\mathit{Lat}(\tilde{v}^{t,k}, \vec{p})$ of each $\tilde{v}^{t,k}$. However, directly measuring these metrics through exhaustive profiling is impractical due to the large number of variants (Challenge~1 in Section~\ref{subsec:challenges}).
To address this, \codename estimates accuracy and latency to accelerate profiling.

The accuracy estimator predicts the accuracy of a stitched variant $A(\tilde{v}^{t,k})$ by using the accuracies of original variants that share its subgraphs. The underlying method is based on the observation that \textit{subgraph-level performance is transferable - subgraphs of high accuracy variants are more likely to produce stitched variants with high accuracy.} Specifically, we first evaluate the accuracy $A(v^{t,i})$ of each original variant $v^{t,i}$ and assign this accuracy to all of its constituent subgraphs, as shown below:
\begin{equation}
X(s_j^{t,i}) = A(v^{t,i}), j \in [1, S]
\label{eq:accuracy1}
\end{equation}
To predict $A(\tilde{v}^{t,k})$, we substitute the subgraph mapping from Equation~\ref{eq:subgraphs_mapping} to Equation~\ref{eq:accuracy1}, which results in:
\begin{equation}
A(\tilde{v}^{t,k}) = A(\{s_j^{t, M[j, i]}\}_{j=1}^S)
\label{eq:accuracy2}
\end{equation}

Based on Equation~\ref{eq:accuracy1} and Equation~\ref{eq:accuracy2}, we build a prediction function $f(\cdot)$ that estimates the accuracy $A(\tilde{v}^{t,k})$ based on the features $X(\{s_j^{t, M[j, i]}\}_{j=1}^S)$. We formulate this estimator as a supervised regression task, with the following objective:
\begin{equation}
\min_{\theta} \sum^{N} \left( A(\tilde{v}^{t,k}) - f_\theta (X(\{s_j^{t, M[j, i]}\}_{j=1}^S)) \right)^2
\label{eq:xgboost_acc}
\end{equation}
where $\theta$ denotes the parameters of $f(\cdot)$ and $N$ is the number of training samples. We implement $f_\theta (.)$ using XGBoost~\cite{chen2016xgboost} trained on a small set of profiled stitched variants.

Regarding latency estimation, we approximate the end-to-end latency of a stitched variant $\tilde{v}^{t,k}$ for the processor placement order $\vec{p}$ by summing the latencies of its subgraphs on the corresponding processors:
\begin{equation}
\mathit{Lat}(\tilde{v}^{t,k}, \vec{p}) \approx \sum_{j=1}^{S} \mathit{Lat}(s_j^{t, M[j, i]}, p_j)
\label{eq:latency_estimation}
\end{equation}
where $\mathit{Lat}(s_j^{t,M[j,i]}, p_j)$ is the measured latency of subgraph $s_j^{t,M[j,i]}$ on processor $p_j$. The inter-subgraph communication cost is not considered, as from empirical evidence it is a relatively low cost (negligible) on shared-memory SoCs since processors access a unified memory space.

\begin{figure}[tp]
    \centering
    \begin{subfigure}[t]{0.49\columnwidth}
        \centering
        \includegraphics[width=\linewidth]{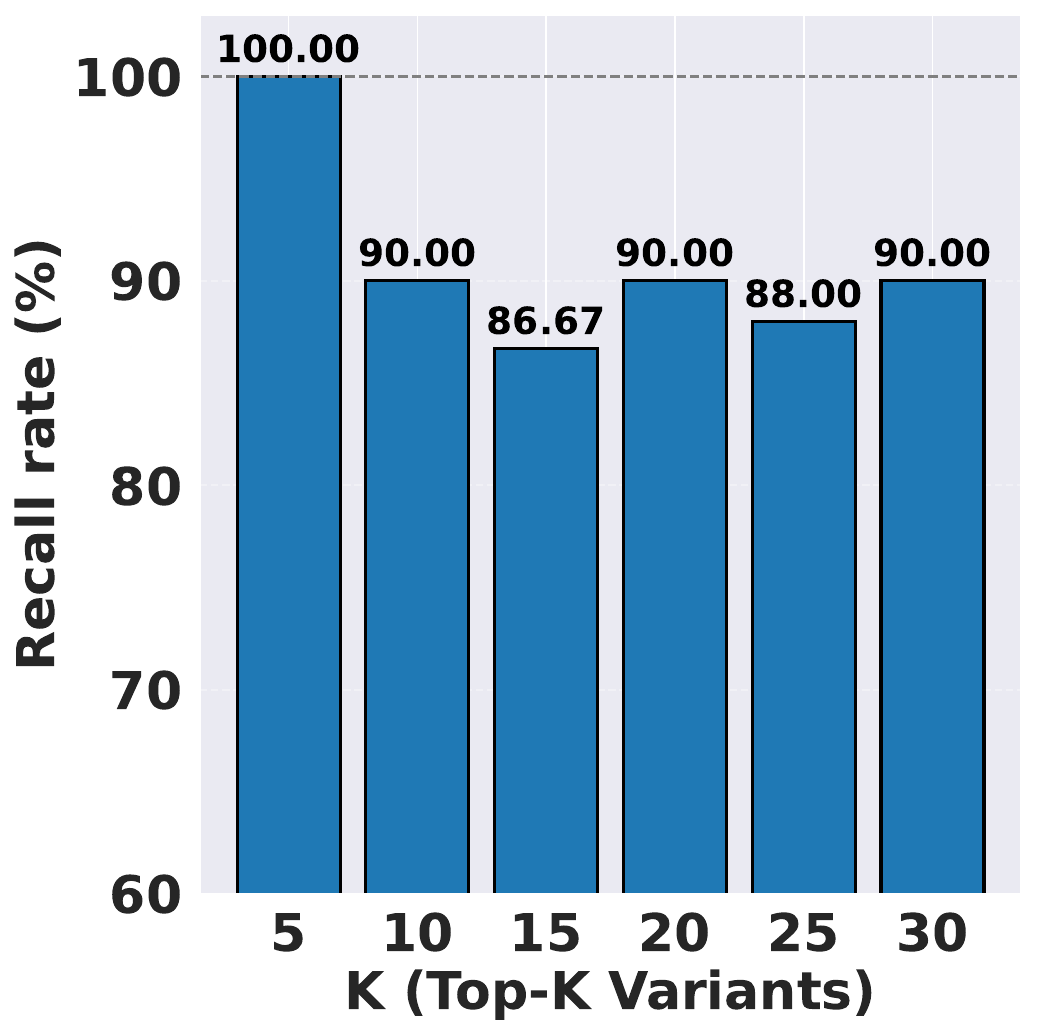}
        \caption{Accuracy Estimator}
        \label{fig:acc_recall}
    \end{subfigure}
    \hfill
    \begin{subfigure}[t]{0.49\columnwidth}
        \centering
        \includegraphics[width=\linewidth]{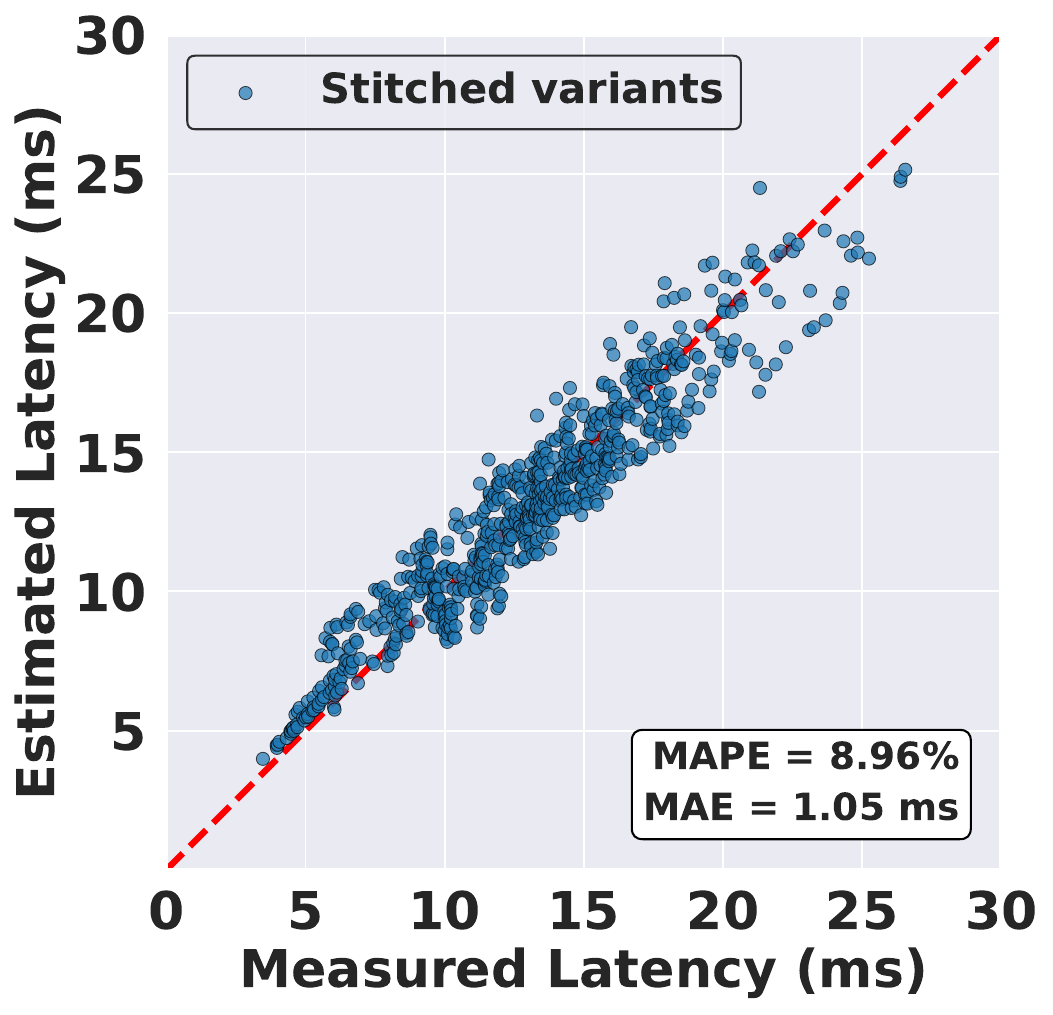}
        \caption{Latency Estimator}
        \label{fig:lat_scatter}
    \end{subfigure}
    \caption{Evaluation of the proposed estimators. (a)~Accuracy estimator: Top-K recall on highest-accuracy variants. (b)~Latency estimator vs. ground-truth.}
\end{figure}

\textbf{How accurate is the \textit{Performance Profiler}?} 
Figure~\ref{fig:acc_recall} presents the evaluation of the accuracy estimator in \codename. We report the Top-K recall rate by using the accuracy estimator to identify stitched variants with top-K accuracies. The results show that \codename can retrieve, on average, 90.78\% of the true top-K variants. Figure~\ref{fig:lat_scatter} shows estimated versus ground-truth latency. The mean absolute error (MAE) is 1.05 ms, while the mean absolute percentage error (MAPE) is 8.9\%. Experimental results confirm the high Top-K recall and low latency error of the proposed estimators.

\textbf{Profiling cost analysis: } 
We further analyze the profiling cost of \codename.
Equation~\ref{eq:total_cost} presents the total profiling cost, which comprises two parts: accuracy estimation for all $V$ variants per task, and latency estimation for each of the $S$ subgraphs on $P$ processors.
\begin{equation}
\mathcal{C}_{\text{\codename}} = 
\underbrace{T \cdot V}_{\text{Accuracy estimation cost}} 
+ 
\underbrace{T \cdot S \cdot V \cdot P}_{\text{Latency estimation cost}}
\label{eq:total_cost}
\end{equation}

\begin{figure}[tp]
    \centering
    \begin{subfigure}[t]{0.49\columnwidth}
        \centering
        \includegraphics[width=\linewidth]{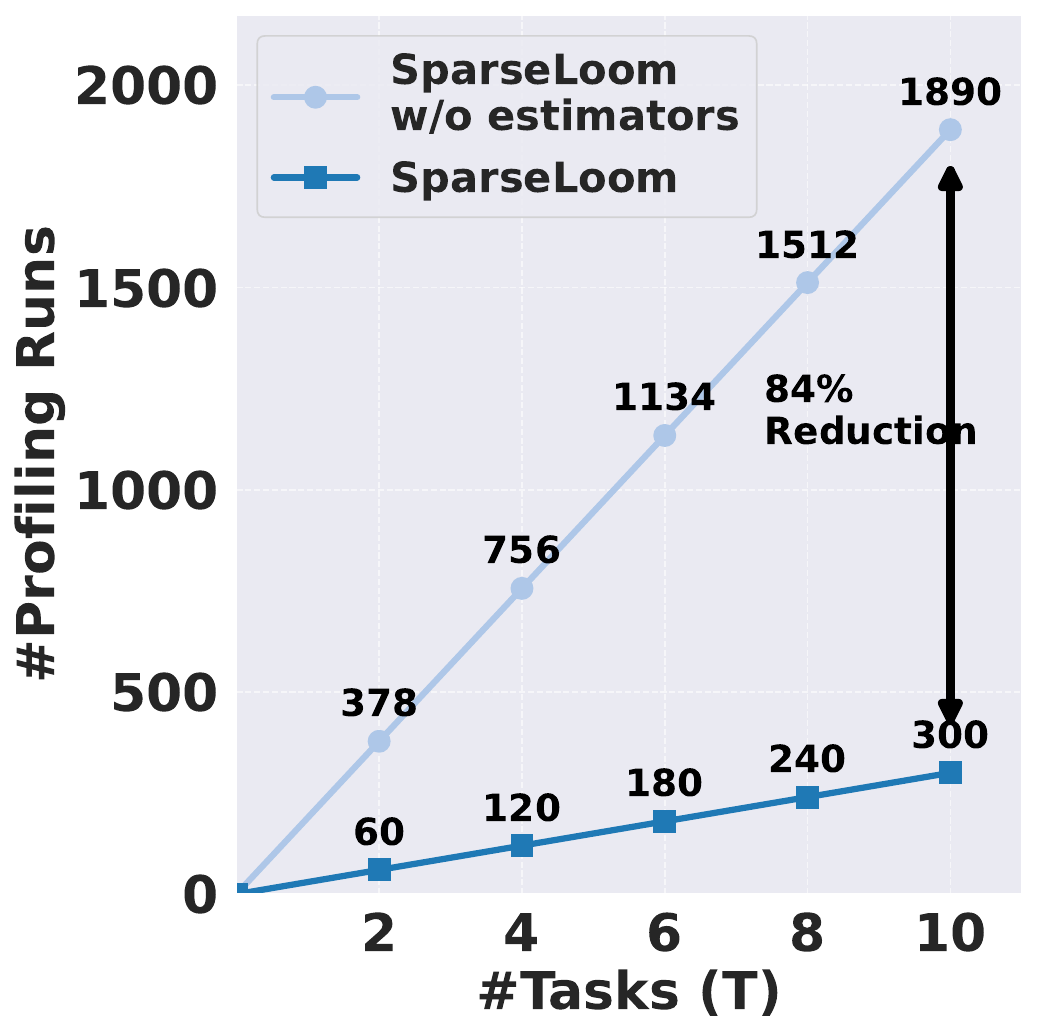}
        \caption{Varying T (P=3, S=3, V=3)}
        \label{fig:profiling_with_T}
    \end{subfigure}
    \hfill
    \begin{subfigure}[t]{0.49\columnwidth}
        \centering
        \includegraphics[width=\linewidth]{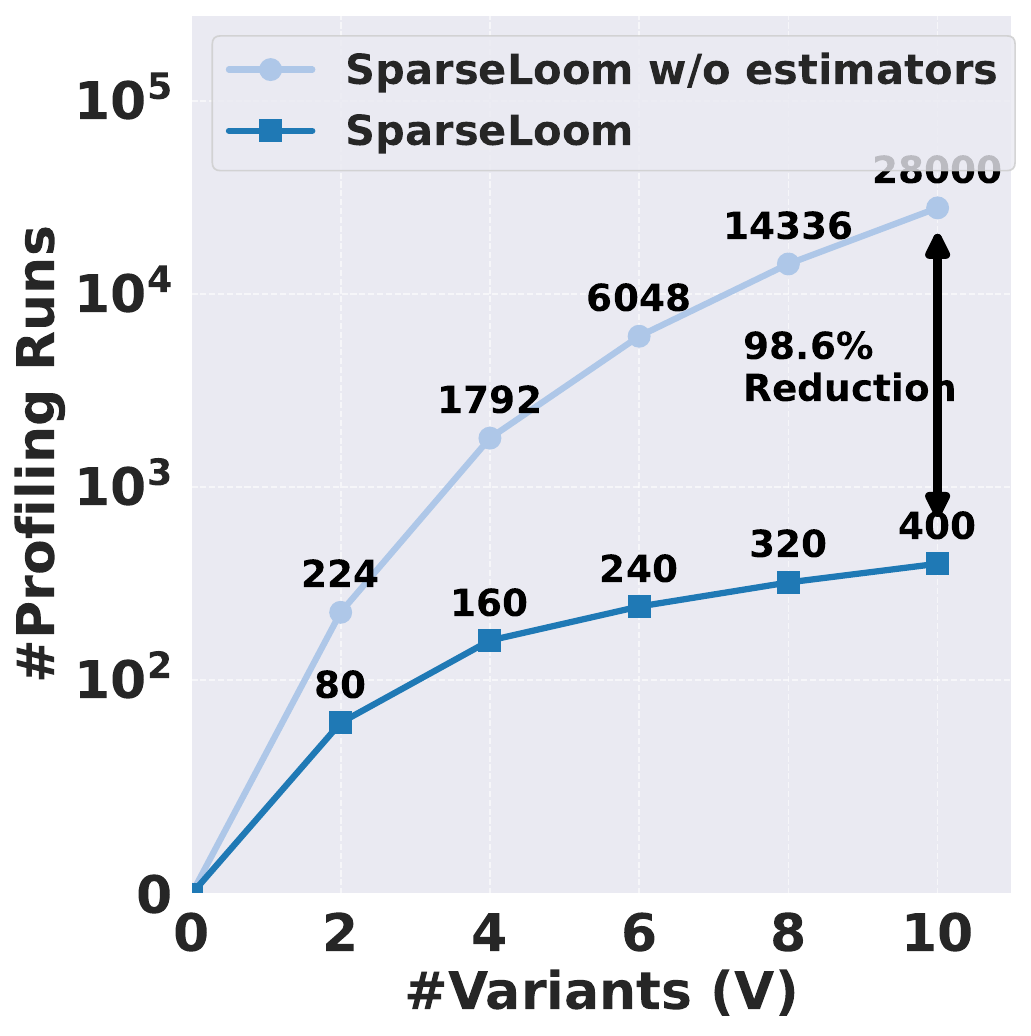}
        \caption{Varying V (P=3, S=3, V=4)}
        \label{fig:profiling_with_V}
    \end{subfigure}
    \caption{Profiling runs of \codename with and without estimators for varying $T$ and $V$.}
    \label{fig:profiling_Optimized}
\end{figure}

Figure~\ref{fig:profiling_Optimized} shows profiling cost of \codename to demonstrate its scalability. When increasing the number of tasks ($T$), \codename reduces the total profiling cost by up to 84\% compared to exhaustive profiling. More importantly, \codename scales linearly with the number of variants per task ($V$), achieving up to 98\% cost reductions. 

\subsection{Optimizing Processor Placement Order and Selecting Stitched Variants}
\label{subsec:optimizer}

Based on the profiling results of stitched variants, \codename then determines a global processor order $\vec{p} = \{p_1, \dots, p_S\}$, shared across all stitched variants, for assigning subgraphs $s_j^{t,i}$ at each position $j$. Adopting a global processor order reduces pipeline bubbles and the overhead of runtime rescheduling~\cite{hetero2pipe2025}. However, existing multi-DNN systems fail to optimize the processor execution order, resulting in sub-optimal latency (Challenge~2 in Section~\ref{subsec:challenges}).

\textbf{Solution in a nutshell:}
\codename jointly optimizes the processor placement order and the final variant selection. We first profile all stitched variants that satisfy the SLO requirement for all possible processor placement orders and identify the best processor order by selecting the one with the lowest average latency across all variants and tasks. Based on this optimal order, we then determine the final variant for each task by choosing the variant with the lowest latency for that order.

Algorithm~\ref{alg:placement} outlines two steps. The optimizer first selects stitched variants that satisfy both accuracy and latency SLOs for each task. The filtered set of stitched variants for task $t$, denoted as $\Theta^{t}$ is the input for subsequent optimization.
Next, the optimizer traverses each candidate processor order $\vec{p} \in \Omega$. For each task $t$, it selects the stitched variant $\tilde{v}^{t,k} \in \Theta^t$ with the lowest latency for $\vec{p}$. The average latency $L(\vec{p})$ across all tasks is then computed, and the processor order that minimizes $L(\vec{p})$ is selected as the optimal placement order $\vec{p}^*$.
Finally, given the selected global placement order $\vec{p}^*$, the optimizer determines the final variant for each task by selecting the stitched variant $\tilde{v}^{t,k} \in \Theta^{t}$ that has the lowest execution latency for $\vec{p}^*$.

The joint optimization of processor placement and variant selection in \codename enables it to determine an optimal processor order that consider different types of sparsity in subgraphs. As a result, the optimizer minimizes the average execution latency across tasks and improves throughput.

\begin{algorithm}[tp]
\caption{Optimizing Processor Placement Order and Selecting Stitched Variants}
\label{alg:placement}
\KwIn{
    $\left\{ SLO^{t}_{acc},\ SLO^{t}_{lat},\ A(\tilde{v}^{t,k}),\ \mathit{Lat}(\tilde{v}^{t,k}, \vec{p}) \right\}_{t=1}^{T}, \mathit{\Omega}$
}
\KwOut{
    $\vec{p}^*$, $\{\tilde{v}^{t,*}\}_{t=1}^T$
}
\tcp{Sparsity-aware processor placement}
\For{$t = 1$ to $T$} {
    $\Theta^{t} = \left\{ \tilde{v}^{t,k} \;\middle|\; A(\tilde{v}^{t,k}) \geq SLO^{t}_{\text{acc}}, \mathit{Lat}(\tilde{v}^{t,k},\vec{p}) \leq SLO^{t}_{\text{lat}}; \exists \vec{p} \in \Omega \right\}$ \\
}

Find $\vec{p}^* \in \mathit{\Omega}$ with minimum $L(\vec{p})$ ;
$L(\vec{p}) = \frac{1}{T} \sum_{t=1}^{T} \min \{ Lat(\tilde{v}^{t,k}, \vec{p}) \}_{ \tilde{v}^{t,k} \in \Theta^{t}}$

\tcp{Final variant selection}

\For{$t = 1$ to $T$}{
    $\tilde{v}^{t,*} = \argmin_{\tilde{v}^{t,k} \in \Theta^{t}} Lat(\tilde{v}^{t,k} | \vec{p}^*)$ \
}
\Return $\vec{p}^*$ and $\{\tilde{v}^{t,*}\}_{t=1}^T$ \
\end{algorithm}

\begin{figure}[tp]
    \centering    \includegraphics[width=1.0\linewidth]{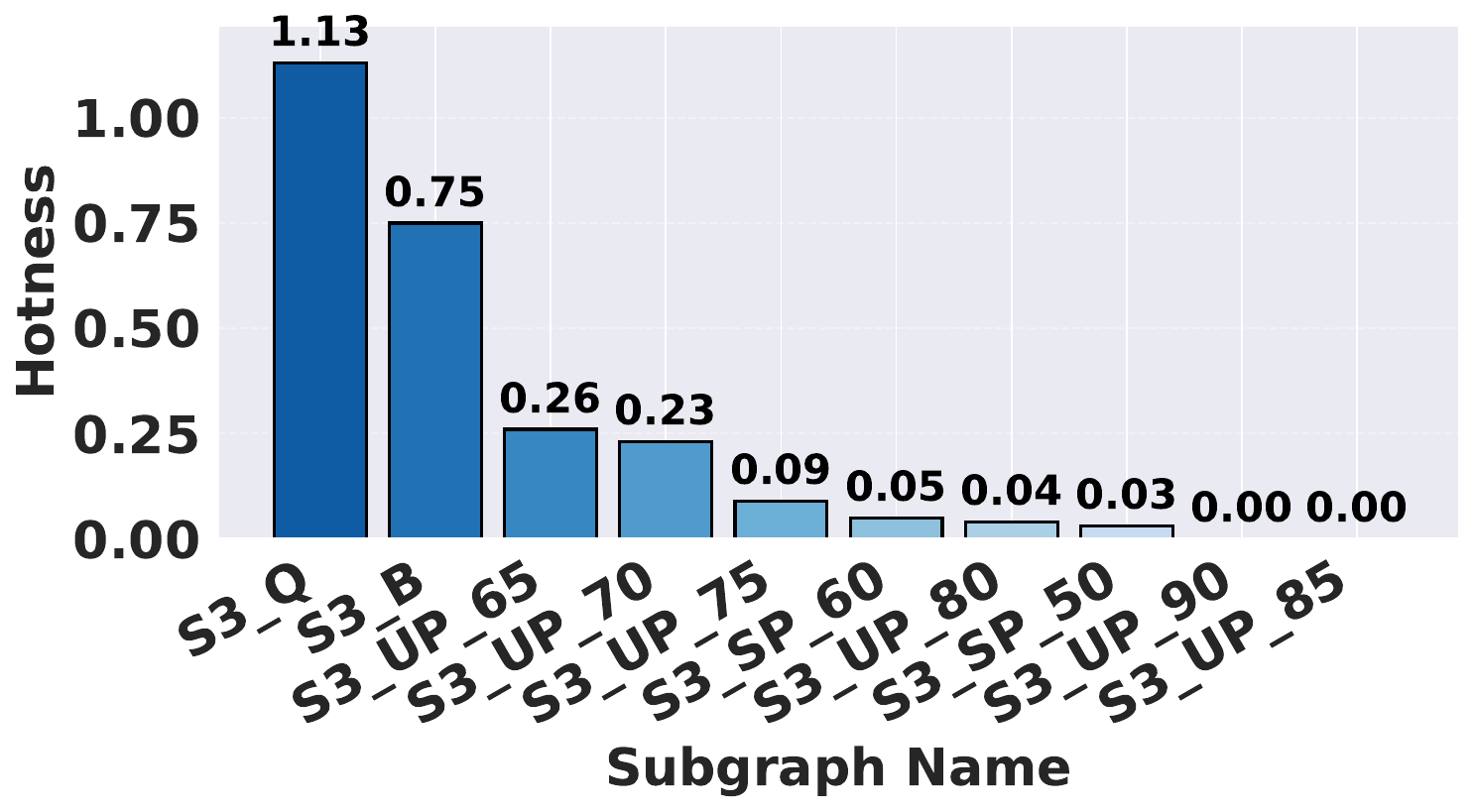}
    \caption{Hotness scores of all subgraphs at the third position, with top four subgraphs are dominant.}
    \label{fig:hotness_score}
\end{figure}

Figure~\ref{fig:hotness_score} presents the hotness scores of all subgraphs that executed at the third position (i.e., $\{s_{j=3}^k\}_{k=1}^{V^S}$). The top four have higher scores, reflecting either their frequent occurrence or distinctiveness. This suggests that preloading the top four subgraphs may satisfy most SLO configurations when memory is limited.

In summary, \codename introduces model stitching to expand the variant space and integrates three key modules to address the challenges of profiling cost, processor placement, and subgraph preloading. Together, these components position \codename as an efficient solution for multi-DNN inference on heterogeneous edge SoCs. 

\begin{algorithm}[tp]
\caption{Hot-Subgraphs Preloading}
\label{alg:preload_selection}
\KwIn{$\left\{ \Theta^t(\sigma) \right\}_{t \in T , \sigma \in \Psi}$, $Mem_{\text{budget}}$; $Mem(s_j^{t,i})$}
\KwOut{$\{\Phi^t\}_{t=1}^T$}

\tcp{Compute hotness scores}
Initialize $H[s_j^{t,i}] \gets 0$ for all $t,i,j$\;
\For{$t = 1$ to $T$}{
  \For{$j = 1$ to $S$}{
    \For{$i = 1$ to $V$}{
      $H[s_j^{t,i}] \gets \sum_{\sigma \in \Psi} 
      \frac{\text{Occur}(s_j^{t,i}, \Theta^t(\sigma))}{|\Theta^t(\sigma)|}$\;
    }
  }
}

\tcp{Greedy preloading under a memory budget}
Initialize $\Phi^t \gets \emptyset$ for all $t$; $M \gets 0$\;
\For{$t = 1$ to $T$}{
  \For{$j = 1$ to $S$}{
    Sort $\{s_j^{t,i}\}_{i=1}^V$ by $H[s_j^{t,i}]$ in descending order\;
    \ForEach{$s_j^{t,i} \in \{s_j^{t,i}\}_{i=1}^V$}{
      \If{$s_j^{t,i} \notin \Phi^t$ \textbf{ and } $M + Mem(s_j^{t,i}) \leq Mem_{\text{budget}}$}{
        $\Phi^t \gets \Phi^t \cup \{s_j^{t,i}\}$\;
        $M \mathrel{+}= Mem(s_j^{t,i})$\;
        \textbf{Break}\;
      }
    }
  }
}
\Return{$\{\Phi^t\}_{t=1}^T$}
\end{algorithm}

\subsection{Preloading Subgraphs of Variants}
\label{subsec:preloader}
Changes to a task’s SLO requirements may necessitate switching the variant at runtime as it may no longer satisfy the updated constraints. To reduce the latency caused by variants switching, preloading is necessary. However, preloading all subgraphs of all variants is memory-intensive (Challenge~3 in Section~\ref{subsec:challenges}). To address this, \codename preloads subgraphs based on a hotness metric. The hotness of a subgraph $s_j^{t,i}$ is defined as
\begin{equation}
\text{Hotness}\left(s_j^{t,i}\right) =
\sum_{\sigma \in \Psi}
\frac{\text{Occur}\left(s_j^{t,i}, \Theta^t(\sigma)\right)}{|\Theta^t(\sigma)|},
\label{eq:hotness}
\end{equation}
where $\Theta^t(\sigma)$ is the set of stitched variants satisfying both accuracy and latency requirements of configuration $\sigma$ for task $t$, and $\Psi$ denotes the set of all SLO configurations with $\sigma \in \Psi$. Here, $\text{Occur}(s_j^{t,i}, \Theta^t(\sigma))$ denotes the number of variants in $\Theta^t(\sigma)$ that contain subgraph $s_j^{t,i}$, and $|\Theta^t(\sigma)|$ is the total number of variants in $\Theta^t(\sigma)$.

The hotness score reflects both the frequency and uniqueness of a subgraph among the variants satisfying different SLOs. \textit{A subgraph achieves a high hotness score either (i) when it is widely reused across variants (high frequency), or (ii) when it is the sole subgraph (or among a few) that can satisfy an SLO configuration (unique).}

Algorithm~\ref{alg:preload_selection} presents the underlying steps of the \textit{Hot-Subgraphs Preloader}.
The hotness score $H[s_j^{t,i}]$ for each subgraph $s_j^{t,i}$ is first computed. Then, for each task $t$, a preloaded subgraph set $\Phi^t$ under the global memory budget $Mem_{\text{budget}}$ is constructed. At each subgraph position $j$, candidate subgraphs are sorted by $H[s_j^{t,i}]$ in descending order, and the preloader greedily loads them into $\Phi^t$ as long as the cumulative memory consumption does not exceed the budget. This strategy ensures that the most frequently reused and critical subgraphs at each position are prioritized for preloading.

\section{Implementation}
\label{sec:implementation}

\codename is implemented in Python using PyTorch~\cite{paszke2019pytorch}. Two settings are supported for deployment on edge SoCs: OpenVINO~\cite{openvino} for Intel SoCs and ONNX Runtime~\cite{onnxruntime} with TensorRT~\cite{tensorrt} for NVIDIA SoCs.

\textbf{OpenVINO on Intel SoCs:}  
OpenVINO is employed as the inference engine on Intel SoCs, as it is specifically optimized for such hardware. OpenVINO supports FP16 and INT8 quantization, as well as acceleration for both unstructured and structured pruning.
To execute a subgraph on Intel SoCs with OpenVINO, \codename compiles the model using the OpenVINO API:
\texttt{Core().compile\_model(model, device\_name)}.
To enable parallel subgraph execution across the CPU, GPU, and NPU, \codename launches independent inference requests in separate threads using OpenVINO's asynchronous interface:
\texttt{infer\_request.start\_async()}.

\textbf{ONNX Runtime with TensorRT on NVIDIA SoCs:}  
For NVIDIA SoCs, \codename adopts ONNX Runtime with TensorRT as inference engines. ONNX Runtime provides a unified interface for executing subgraphs on both the GPU and ARM CPU. TensorRT enables high performance GPU execution with support for low precision computation (e.g., FP16 and INT8) and structured pruning. 
To run subgraphs on NVIDIA SoCs, \codename first exports each subgraph as an ONNX model. For GPU execution, it instantiates an \texttt{InferenceSession} with either \texttt{TensorrtExecutionProvid\\er} or \texttt{CUDAExecutionProvider}. For CPU execution, \texttt{CPUExec\\utionProvider} is used.
Subgraph inference is performed via \texttt{ort.InferenceSession(model, providers)} within parallel threads to enable concurrent execution.

\section{Evaluation}
\label{sec:evaluation}

In this section, we evaluate \codename across three edge platforms and four dataset–model pairs, comparing it against six baselines. The main observations are: 
\begin{itemize}[leftmargin=*]
    \item \textbf{End-to-end performance gain}: \codename reduces SLO violation rates by up to 74\% and improves throughput by up to 2.31$\times$ compared to other baselines.
    \item \textbf{Individual modules are effective}: \codename reduces profiling time by up to 99\% with efficient estimators, improves throughput by up to 2$\times$ via optimized processor placement, and lowers memory cost by an average of 28\% without compromising SLO violation rates.
\end{itemize}

\subsection{Experimental Setup}
\label{subsec:setup}

\textbf{Platforms:}
Table~\ref{table:devices} presents the three platforms used in our evaluation.
They represent a spectrum of edge SoCs, ranging from high-end desktops to low-power embedded devices. 

\textbf{Tasks, datasets and base models:} 
Table~\ref{tab:dnn-models} shows the four tasks with their corresponding datasets and models. We cover diverse tasks across vision, language, activity recognition, and audio, each paired with representative datasets and base models. These task, dataset, and models are commonly used at the edge for on-device ML~\cite{hetero2pipe2025,karatzas2023omniboost}.

\textbf{Sparse model zoo:}
For each task, we build a sparse model zoo of ten variants commonly adopted in the literature~\cite{taufique2024tango,sen2024elastically,fang2018nestdnn}, including one dense base model and nine sparse variants. However, \codename does not require a fixed number of variants, as it exponentially expands the variant space via model stitching rather than relying on pre-generated variants.
The base models are sourced from public model repositories (e.g., Hugging Face), and the corresponding sparse variants are pre-generated using model compression frameworks such as OpenVINO’s Neural Network Compression Framework (NNCF)~\cite{nncf} and ONNX Runtime~\cite{onnxruntime}.
Appendix~\ref{appendix:variants} provides a complete list of all variants and details their preparation and software/hardware support required.

\textbf{Metrics:} The measured metrics are:  
\textit{SLO Violation Rate} -  We define the SLO violation rate as the fraction of tasks that fail to meet either latency or accuracy SLOs. To comprehensively evaluate this, we report the average violation rate over all task arrival combinations. A task arrival combination refers to a particular order of all tasks. For the four tasks used in our experiments, there are 24 such combinations in total. \textit{Throughput} - the number of completed inference queries per unit time. Specifically, we repeatedly run four tasks for 100 queries (denoted as one run) with each task using a batch size of 1, and compute throughput as
$\frac{N}{T_{\text{total}}}$,  
where $N$ is the total number of completed inference queries and $T_{\text{total}}$ is the total execution time. All results are averaged over 10 runs\footnote{Error bars are not included as a low variance is observed across runs.}.

\begin{table*}[ht]
\centering
\caption{Platforms used in the evaluation.}
\begin{tabular}{P{5.9cm} P{2.9cm} P{4.9cm} P{2.7cm}}
\Xhline{2\arrayrulewidth}
\textbf{Platform} & \textbf{CPU} & \textbf{GPU} & \textbf{NPU} \\
\Xhline{2\arrayrulewidth}
Desktop (Intel Core Ultra 7 265K) & \makecell[c]{x86-64; 20-core}  &  4-$\text{X}^{e}$core  &  Intel AI Boost   \\
\hline
Laptop  (Intel Core Ultra 5 135U) & \makecell[c]{x86-64; 12-core}   & 4-$\text{X}^{e}$core   &  Intel AI Boost   \\
\hline
NVIDIA Jetson AGX Orin (MAXN mode)
&  
ARM Cortex 12-core
& 
2048-core Ampere; 64 Tensor cores
& \makecell[c]{ N/A}
\\
\Xhline{2\arrayrulewidth}
\end{tabular}
\label{table:devices}
\end{table*}

\begin{table}[t]
  \centering
  \caption{Tasks, datasets, and models for evaluation. The selected tasks and models are representative of the AR use case discussed in Section \ref{sec:introduction}.}
  \label{tab:dnn-models}
  \setlength{\tabcolsep}{1pt}
  \begin{tabular}{P{3.6cm} P{2.7cm} P{2.1cm}}
    \Xhline{2\arrayrulewidth}
    \textbf{Task type} & \textbf{Dataset} & \textbf{Base Model} \\
    \Xhline{2\arrayrulewidth}
    Image classification       & \small{ImageNet-1K~\cite{deng2009imagenet}}      & \small{ResNet-101~\cite{he2016deep}} \\
    \hline
    Sentiment classification   & SST-2~\cite{socher2013recursive}             & \small{BERT-Base~\cite{devlin2019bert}} \\
    \hline
    \small{Human activity recognition} & HAR~\cite{nagadia2021har}               & ViT-Small~\cite{dosovitskiy2020image} \\
    \hline
    Speech recognition & \small{LibriSpeech ASR~\cite{panayotov2015librispeech}} & Wav2vec2~\cite{baevski2020wav2vec} \\
    \Xhline{2\arrayrulewidth}
  \end{tabular}
\end{table} 

\textbf{Baseline design:}  
To enable a comprehensive evaluation of \codename, we categorize existing multi-DNN inference systems along two dimensions. The first dimension is the \textit{variant selection strategy}. Systems may adopt a single variant per task (SV), either accuracy-optimal (AO) or latency-optimal (LO), or support adaptive variants per task (AV) that adjust to different SLO configurations. Representative examples include Pipe-it~\cite{wang2019high}, Pantheon~\cite{han2024pantheon}, and RT-mDL~\cite{ling2021rt} for SV-AO; Hetero$^2$Pipe~\cite{hetero2pipe2025}, Band~\cite{jeong2022band}, and OmniBoost~\cite{karatzas2023omniboost} for SV-LO; and Tango~\cite{taufique2024tango}, ESIM~\cite{sen2024elastically}, and NestDNN~\cite{fang2018nestdnn} for AV.  
The second dimension is whether the system employs \textit{variant partitioning}. Partitioned (P) systems decompose a variant into subgraphs for concurrent execution on processors~\cite{hetero2pipe2025,jeong2022band,ling2022blastnet}, whereas non-partitioned (NP) systems treat each variant as a monolithic unit~\cite{wang2019high,han2024pantheon,han2022microsecond}.  
    
The two dimensions collectively yield six baselines: \textbf{SV-AO-P}, \textbf{SV-AO-NP}, \textbf{SV-LO-P}, \textbf{SV-LO-NP}, \textbf{AV-P}, and \textbf{AV-NP}, which covers the features space of state-of-the-art multi-DNN inference systems. \codename is positioned in the adaptive-variant with partitioning (AV-P) category. However, it advances beyond prior systems by \textit{model stitching}.

\textbf{SLO configurations:}
To evaluate \codename and other baselines under diverse SLOs, we systematically construct a set of accuracy–latency configurations. For each task, we first measure the accuracy and latency of all sparse variants and record the observed accuracy and latency ranges. We then extend these ranges: the latency range is extended by 20\% (from 80\% of the minimum latency to 120\% of the maximum latency), and the accuracy range is extended by $\pm$2\% (from 2\% below the minimum accuracy to 2\% above the maximum accuracy). Based on these extended ranges, we uniformly sample five points from the accuracy range and five points from the latency range. The Cartesian product of these two sets yields $5 \times 5 = 25$ SLO configurations.

For instance, if the observed accuracy range across all variants is [85\%, 92\%] and the latency range is [50\,ms, 120\,ms], the extended ranges become [83\%, 94\%] for accuracy and [40\,ms, 144\,ms] for latency. We then uniformly sample five accuracy points, e.g., \{83\%, 85.75\%, 88.5\%, 91.25\%, 94\%\}, and five latency points, e.g., \{40\,ms, 66\,ms, 92\,ms, 118\,ms, 144\,ms\}. 
The final SLO configurations are given by the Cartesian product of these two sets, yielding 25 accuracy–latency pairs.

\begin{figure*}[ht]
  \centering
  \begin{subfigure}[t]{0.32\linewidth}
    \centering
    \includegraphics[width=\linewidth]{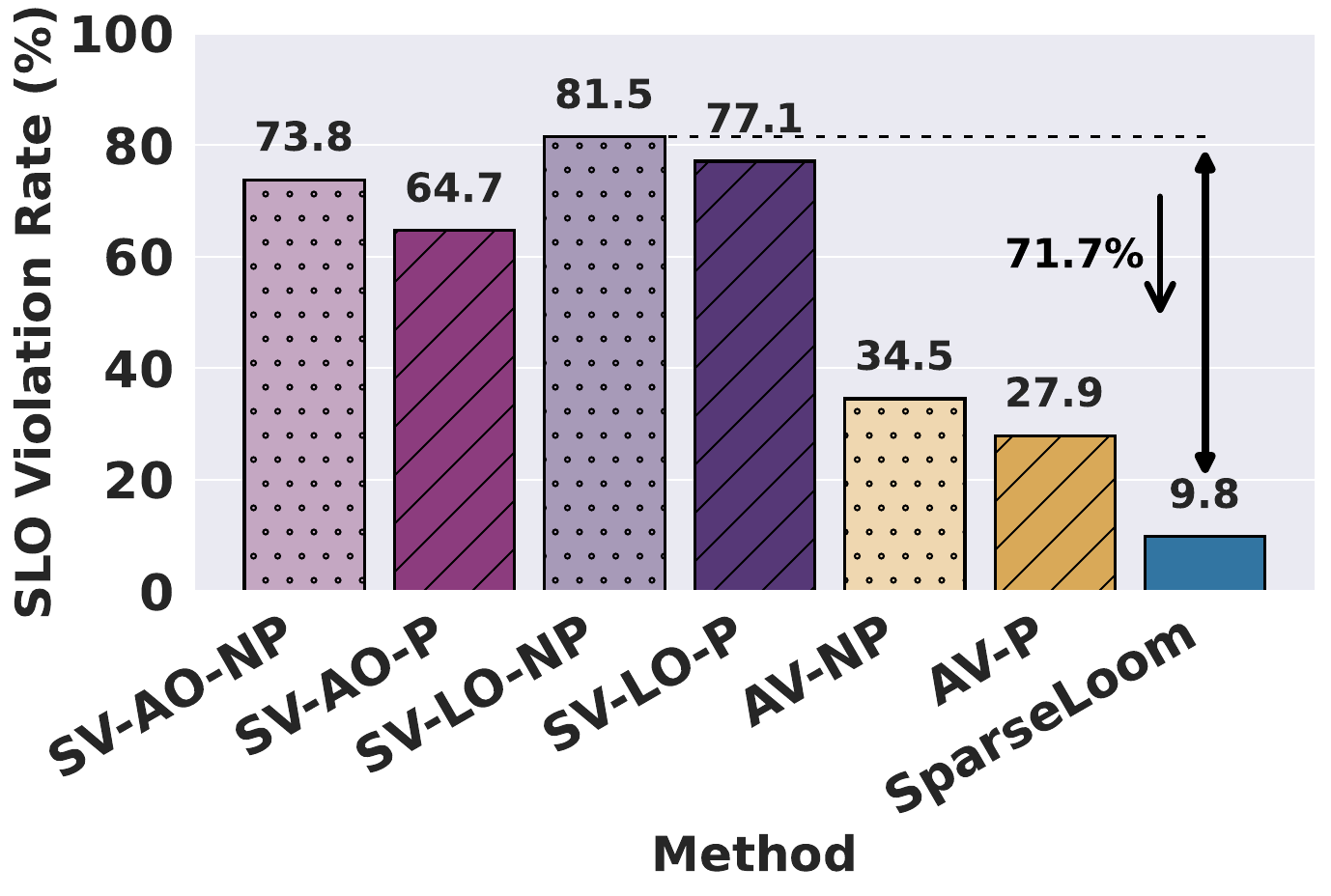}
    \caption{Desktop}
    \label{fig:slo_desktop}
  \end{subfigure}
  \hfill
  \begin{subfigure}[t]{0.32\linewidth}
    \centering
    \includegraphics[width=\linewidth]{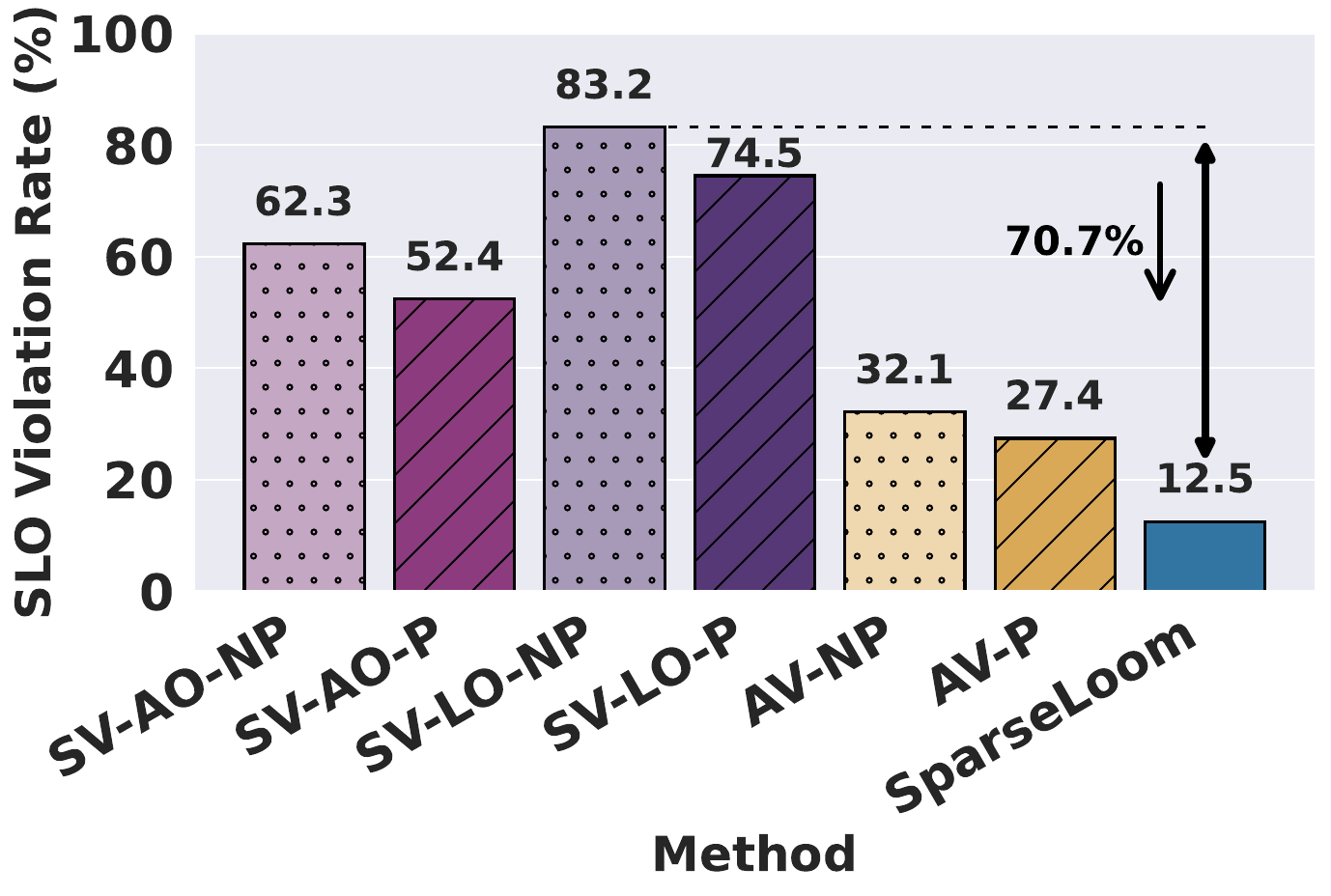}
    \caption{Laptop}
    \label{fig:slo_laptop}
  \end{subfigure}
  \hfill
  \begin{subfigure}[t]{0.32\linewidth}
    \centering
    \includegraphics[width=\linewidth]{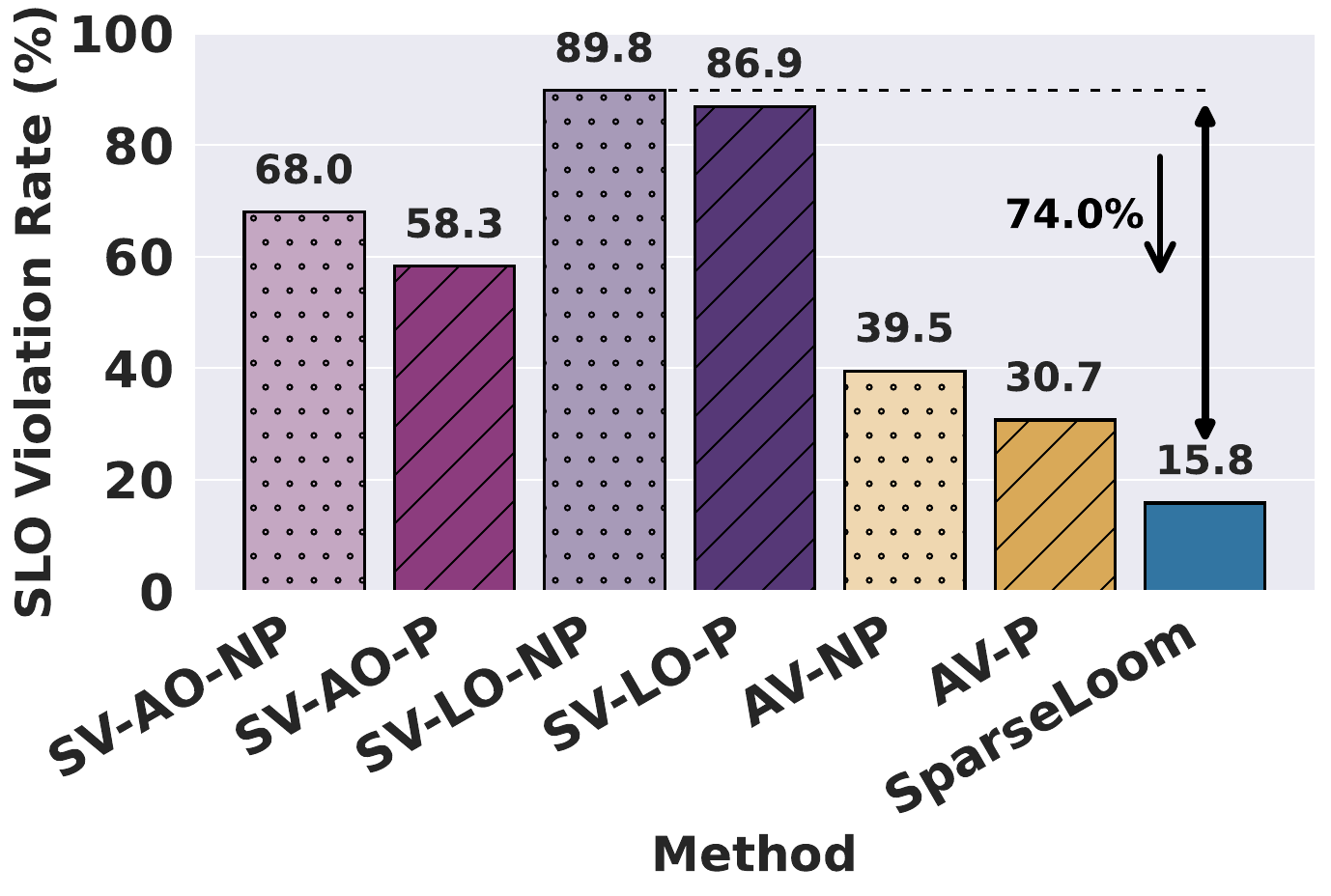}
    \caption{Jetson Orin}
    \label{fig:slo_jetson}
  \end{subfigure}
  \caption{SLO violation rates of \codename and baselines across three SoCs. \codename consistently achieves lower violation rates than all baselines, with reductions of up to 74\%.}
  \label{fig:multi_violation}
\end{figure*}

\begin{figure*}[ht]
  \centering
  \begin{subfigure}[t]{0.32\linewidth}
    \centering
    \includegraphics[width=\linewidth]{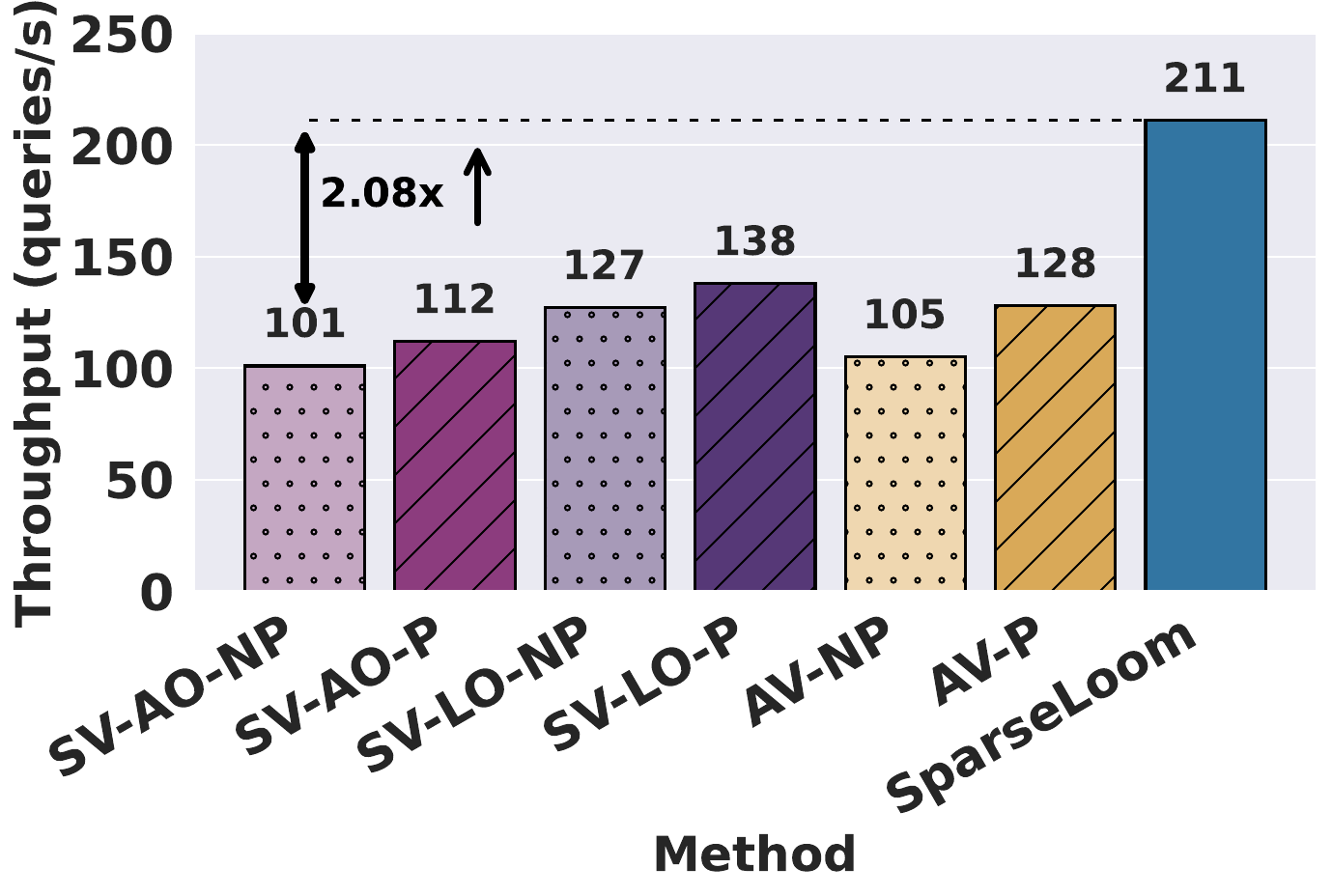}
    \caption{Desktop}
    \label{fig:throughput_desktop}
  \end{subfigure}
  \hfill
  \begin{subfigure}[t]{0.32\linewidth}
    \centering
    \includegraphics[width=\linewidth]{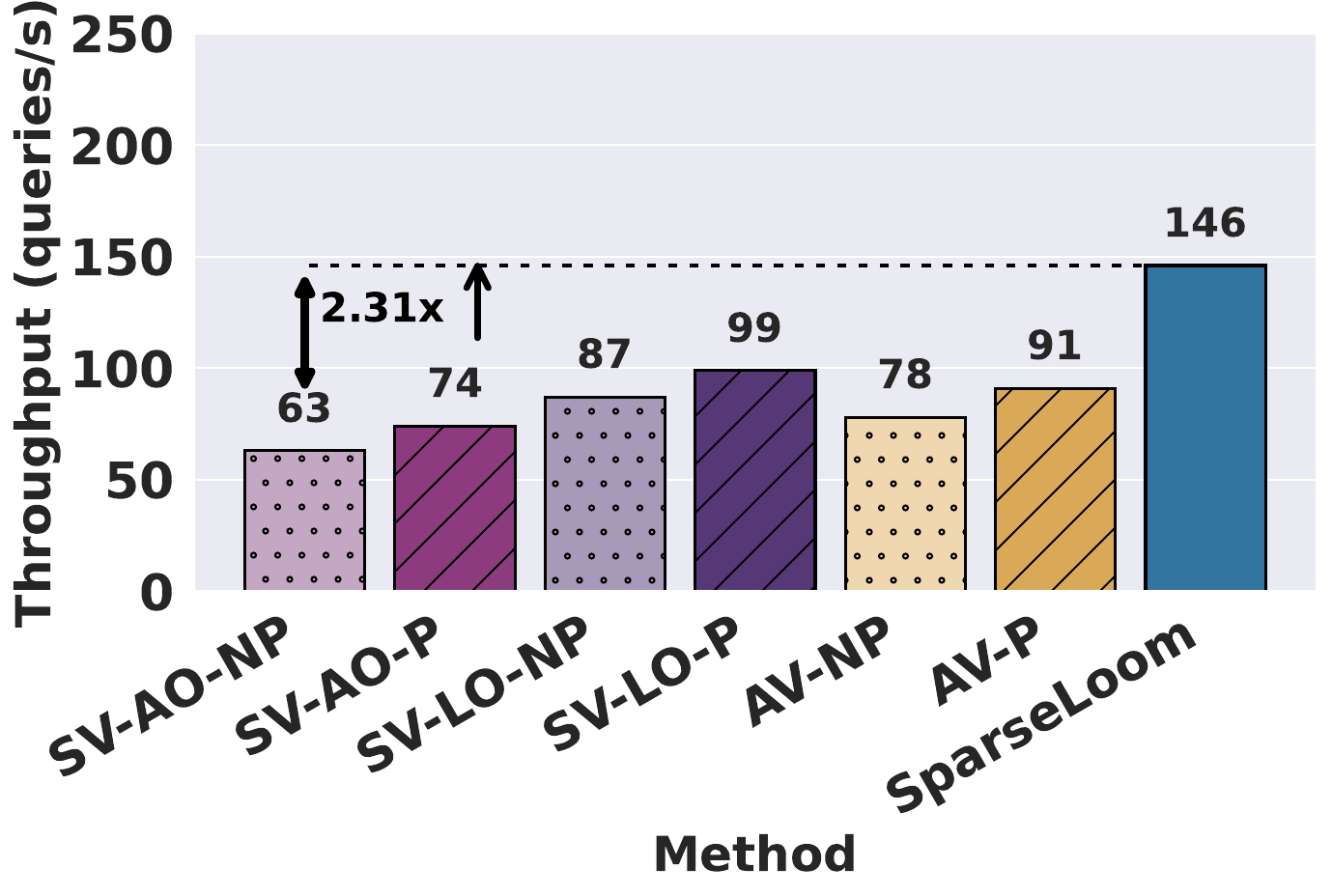}
    \caption{Laptop}
    \label{fig:throughput_laptop}
  \end{subfigure}
  \hfill
  \begin{subfigure}[t]{0.32\linewidth}
    \centering
    \includegraphics[width=\linewidth]{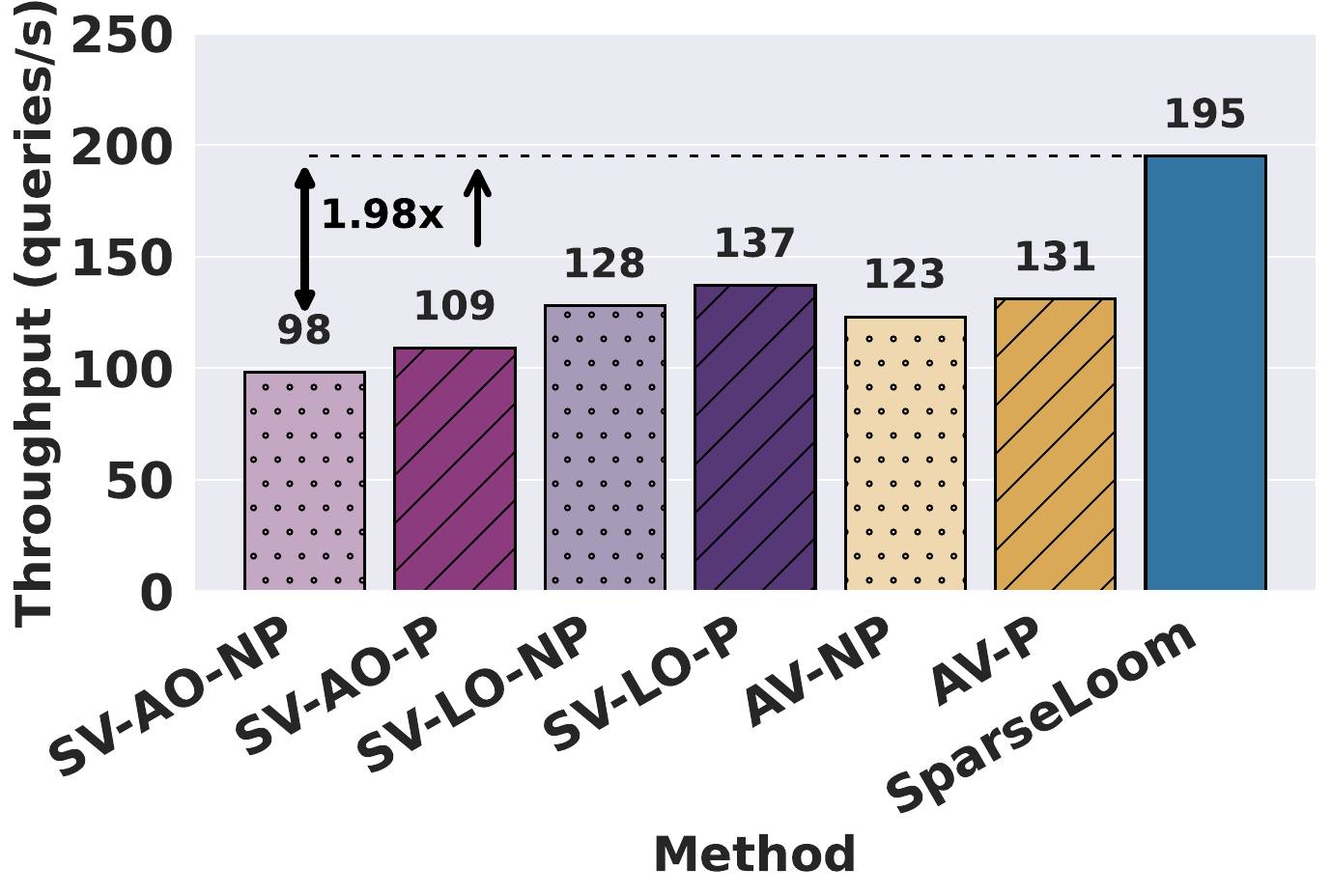}
    \caption{Jetson Orin}
    \label{fig:throughput_jetson}
  \end{subfigure}
  \caption{Inference throughput of \codename and baselines across three SoCs. \codename consistently achieves higher throughput than all baselines, with improvements of up to 2.31$\times$.}
  \label{fig:multi_throughput}
\end{figure*}

\subsection{End-to-end Performance}

\textbf{Lower SLO violation rate:}
Figure~\ref{fig:multi_violation} shows the SLO violation rates of \codename and other baselines across three testbeds running four tasks in parallel. Among all baselines, SV-LO methods (SV-LO-P and SV-LO-NP) show the highest violation rates, as they adopt a single variant per task optimized only for latency. By contrast, adaptive variant selection from sparse variants significantly reduces violation rates. On all three testbeds, both \codename and AV methods (AV-P and AV-NP) achieve much lower violation rates than SV methods (including SV-AO-P, SV-AO-NP, SV-LO-P, and SV-LO-NP). Moreover, \codename further improves upon other AV methods, highlighting the benefits of incorporating model stitching into adaptive-variant systems. Specifically, compared to SV methods, \codename reduces violation rates by up to 74\% (on Jetson Orin against SV-LO-NP), and compared to AV methods, it achieves up to 24.7\% reduction (on desktop against AV-NP). Overall, \codename consistently delivers reduced SLO violations.

We further evaluate \codename under \textit{accuracy- and latency-guaranteed SLOs}, where the accuracy requirement is fixed to the highest value across all variants or the latency requirement to the lowest value, respectively (see Appendix~\ref{appendix:experiments_slo} for details). \codename consistently reduces SLO violation rates by up to 73.6\% under accuracy-guaranteed SLOs and 68.2\% under latency-guaranteed SLOs, demonstrating its effectiveness even in extreme scenarios where no accuracy or latency compromise is allowed.

\textbf{Higher inference throughput:} 
Figure~\ref{fig:multi_throughput} shows the inference throughput of \codename and other baselines across three testbeds. Partitioning methods (SV-AO-P, SV-LO-P, and AV-P) achieve higher throughput than their non-partitioning counterparts (SV-AO-NP, SV-LO-NP, and AV-NP), owing to the benefits of parallel execution of subgraphs across multiple processors. Overall, SV and AV methods deliver comparable throughput; even SV-LO-P, which greedily selects the variant with the lowest latency, yields marginal improvements over other baselines. In contrast, \codename achieves substantially higher throughput—improving by up to 2.31$\times$ on the laptop compared to SV-AO-NP, and by up to 1.53$\times$ over the best baseline (SV-LO-P) on the desktop. These gains arise from model stitching, which offers more stitched variants and optimized placement orders (see Section~\ref{subsec:individual modules}).

In summary, \codename consistently outperforms state-of-the-art baselines by achieving both lower SLO violation rates and higher inference throughput. It reduces SLO violations by up to 74\% and improves throughput by up to 2.31$\times$ across diverse edge SoCs.

\begin{figure*}[t!]
  \centering
  \begin{subfigure}[t]{0.32\linewidth}
    \centering
    \includegraphics[width=\linewidth]{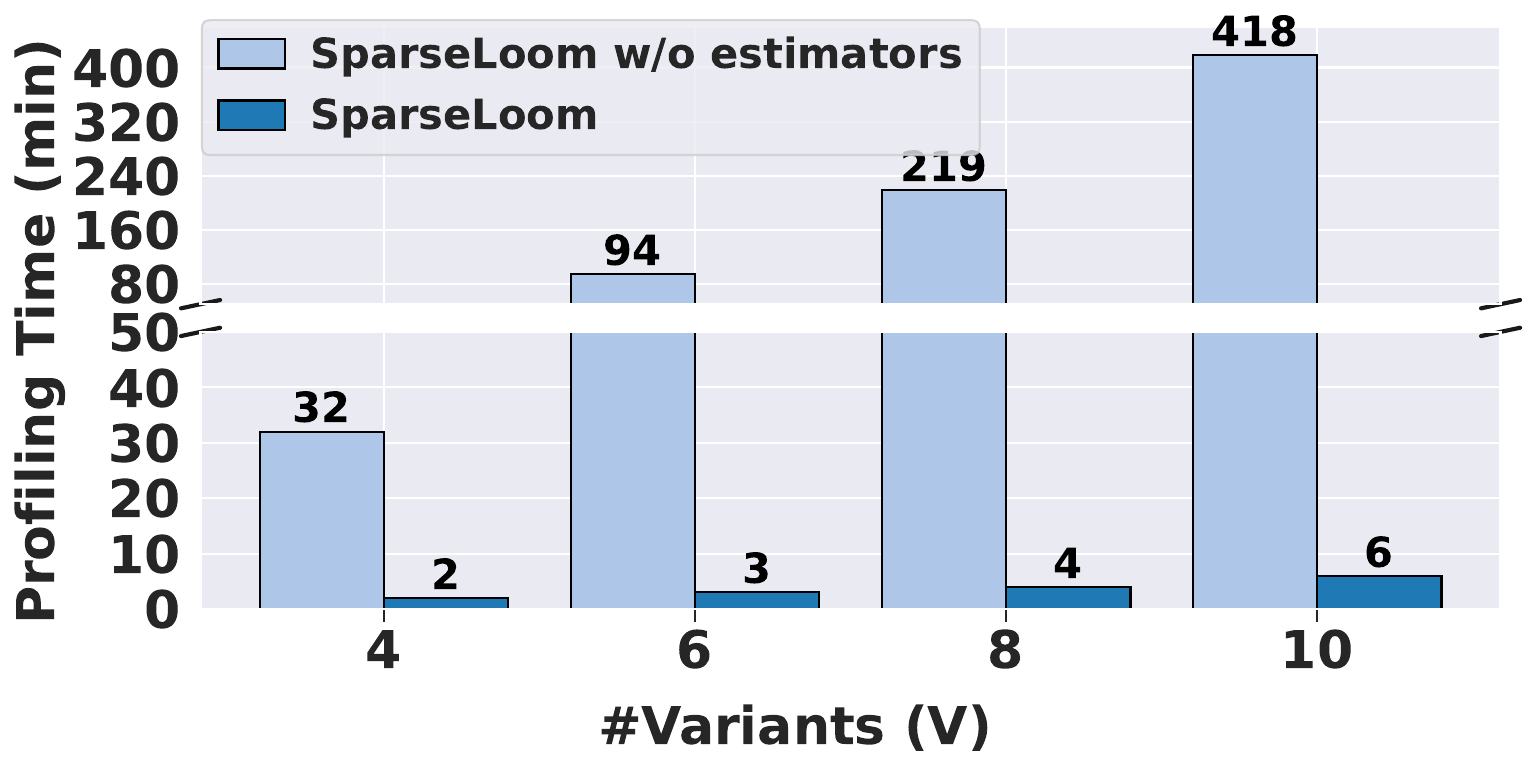}
    \caption{Desktop}
    \label{fig:Profiling_cost_desktop}
  \end{subfigure}
  \hfill
  \begin{subfigure}[t]{0.32\linewidth}
    \centering
    \includegraphics[width=\linewidth]{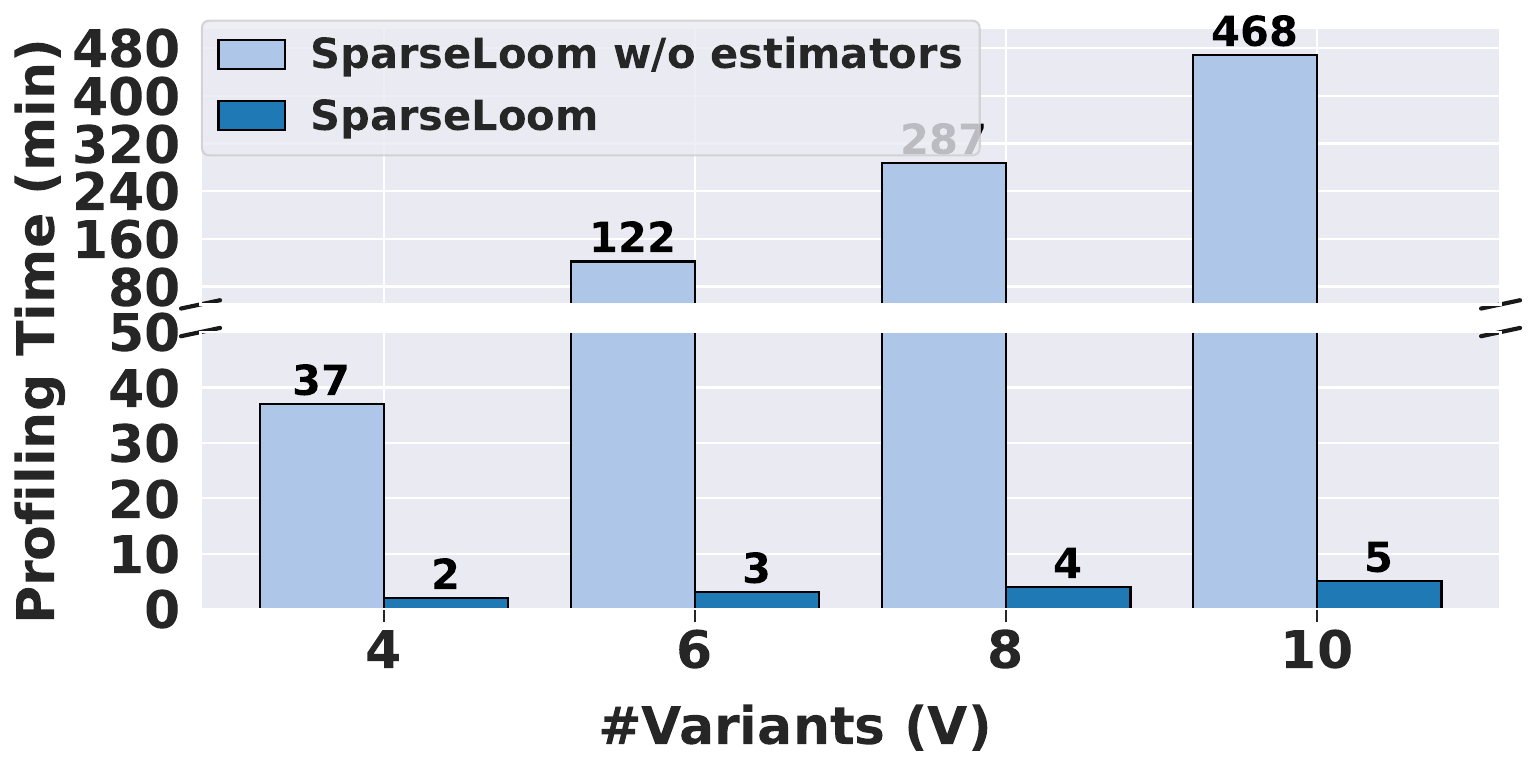}
    \caption{Laptop}
    \label{fig:Profiling_cost_laptop}
  \end{subfigure}
  \hfill
  \begin{subfigure}[t]{0.32\linewidth}
    \centering
    \includegraphics[width=\linewidth]{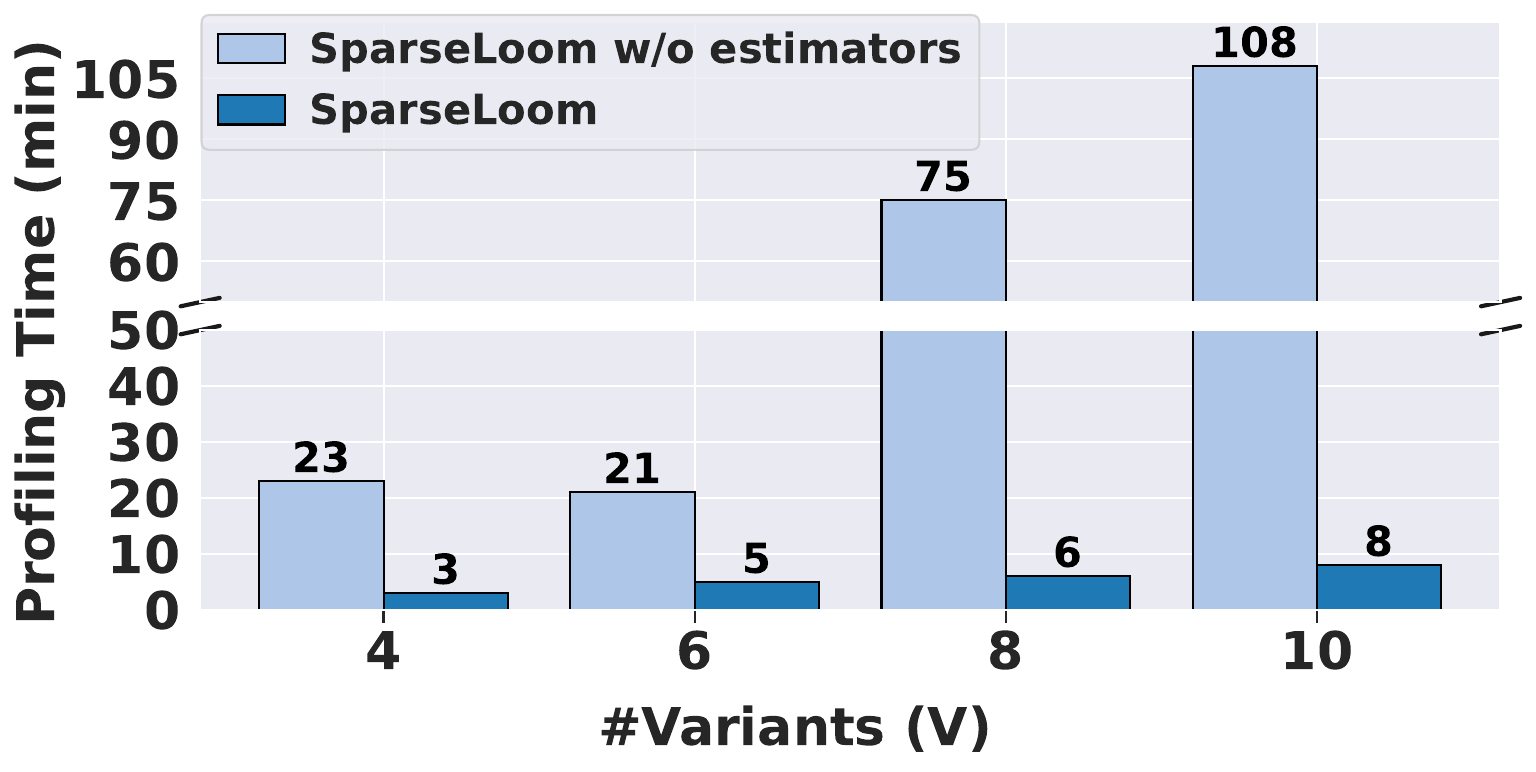}
    \caption{Jetson Orin}
    \label{fig:Profiling_cost_jetson}
  \end{subfigure}
  \caption{Profiling time (minutes) for \codename with and without estimators. \codename reduces profiling cost to under 8 minutes (up to 99\% reduction) on all SoCs.}
  \label{fig:Profiling_cost_eval}
\end{figure*}

\begin{figure*}[t!]
  \centering
  \begin{subfigure}[t]{0.32\linewidth}
    \centering
    \includegraphics[width=\linewidth]{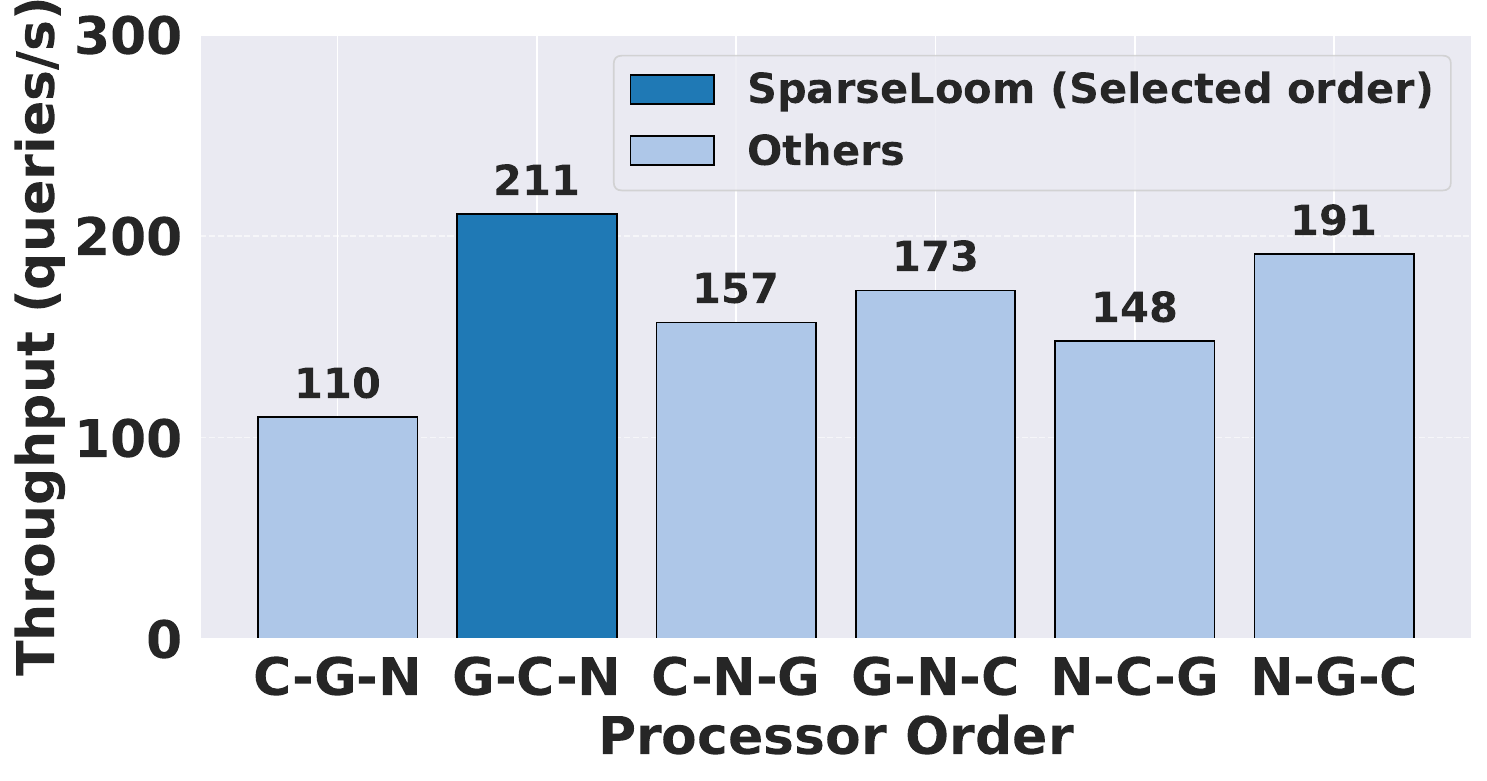}
    \caption{Desktop}
    \label{fig:placement_order_throughput_Desktop}
  \end{subfigure}
  \hfill
  \begin{subfigure}[t]{0.32\linewidth}
    \centering
    \includegraphics[width=\linewidth]{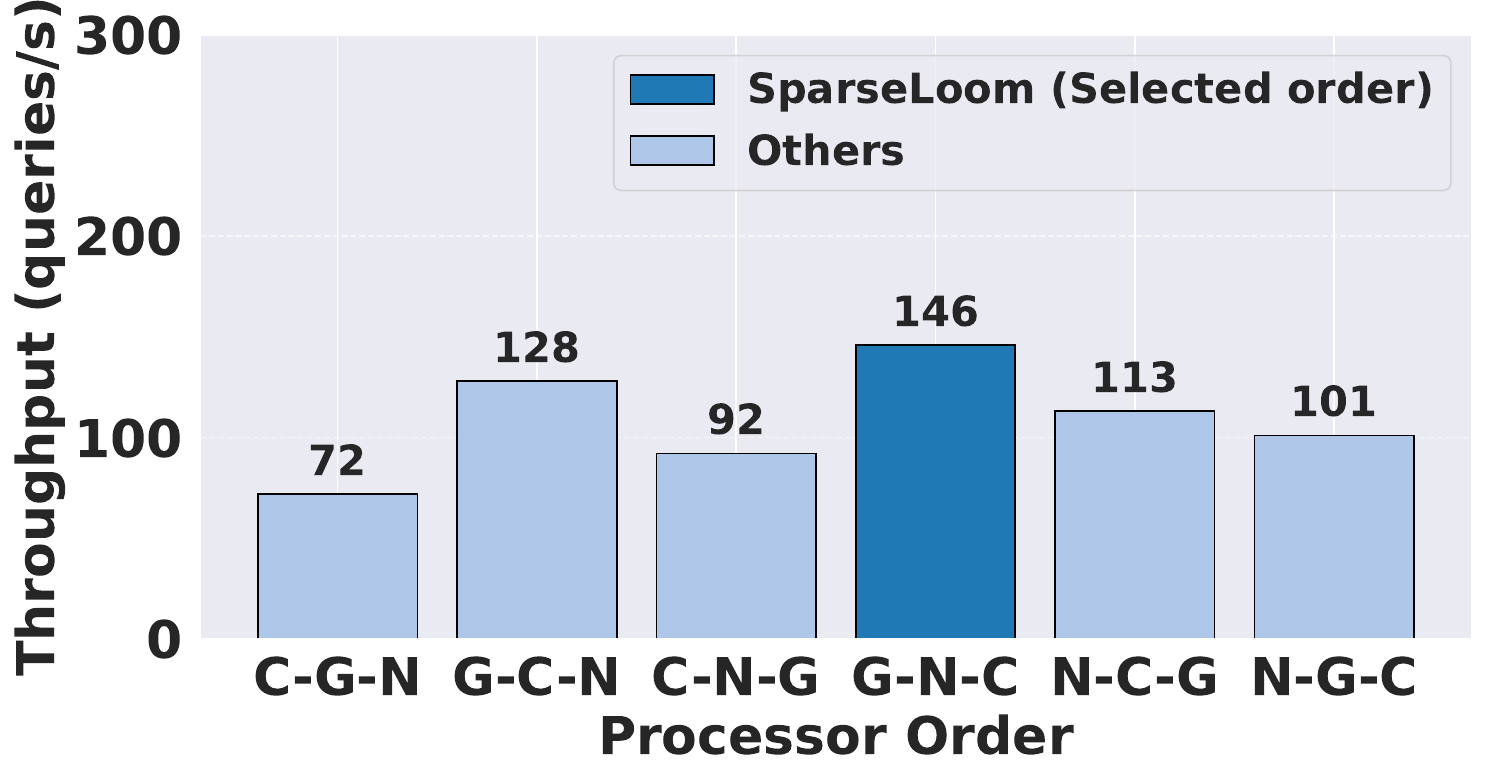}
    \caption{Laptop}
    \label{fig:placement_order_throughput_laptop}
  \end{subfigure}
  \hfill
  \begin{subfigure}[t]{0.32\linewidth}
    \centering
    \includegraphics[width=\linewidth]{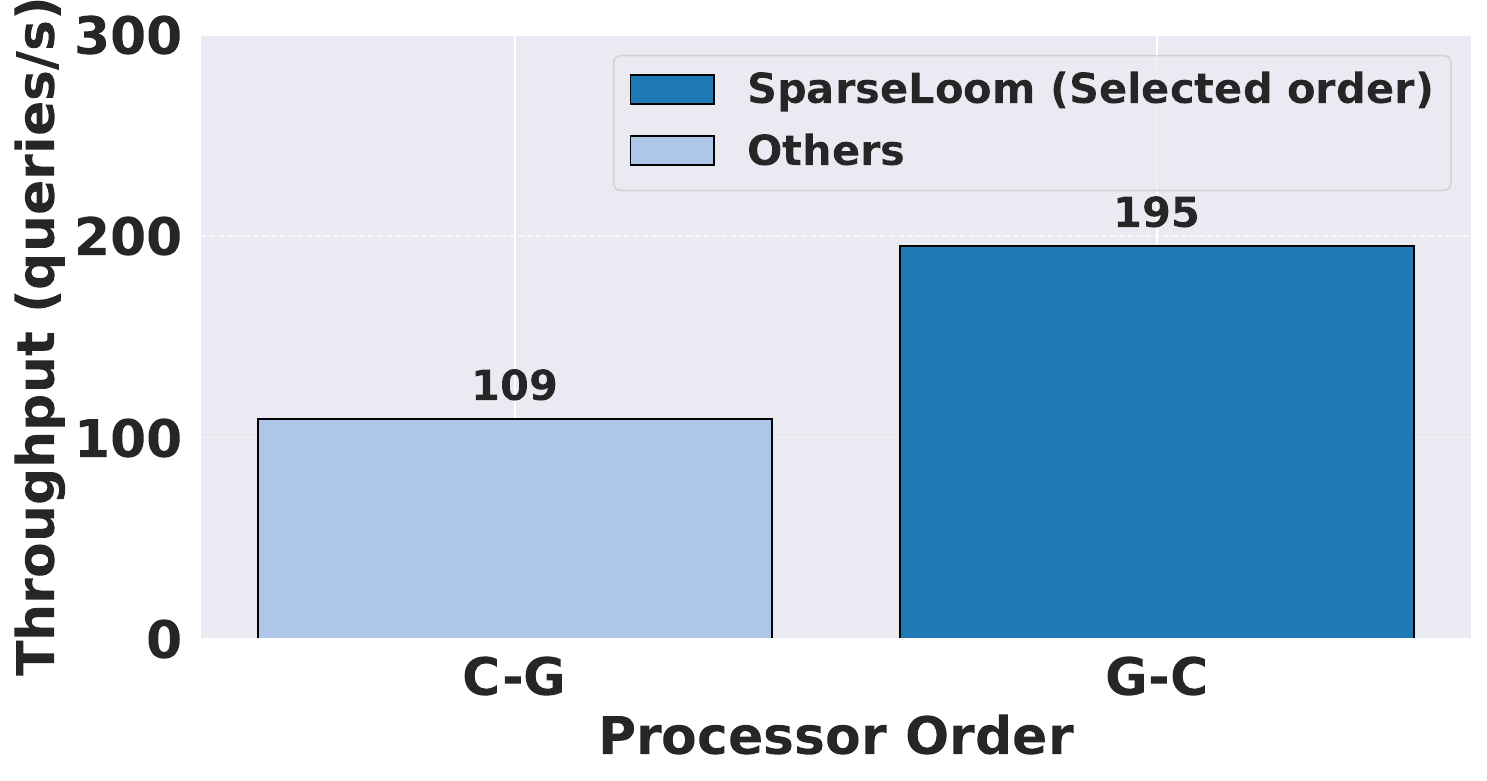}
    \caption{Jetson Orin}
    \label{fig:placement_order_throughput_jetson}
  \end{subfigure}
  \caption{Impact of processor placement order on inference throughput (C = CPU, G = GPU, N = NPU). \codename automatically selects the best placement, achieving up to $2\times$ higher throughput.}
  \label{fig:placement_order_throughput}
\end{figure*}

\subsection{Evaluation of Individual Modules}
\label{subsec:individual modules}
In this section, we further evaluate the three main modules of \codename. Specifically, we examine: (i) profiling time for accuracy and latency with and without estimators, (ii) the impact of processor placement optimization on throughput, and (iii) the effect of subgraph preloading on SLO violations under varying memory budgets.

\textbf{Profiling overhead of \codename:}
Figure~\ref{fig:Profiling_cost_eval} compares the profiling times of \codename with and without accuracy and latency estimators across all three testbeds. As the number of variants increases, the profiling overhead without estimators grows exponentially. For instance, with 10 variants (the setting used in our experiments), profiling without estimators on the laptop takes about 468 minutes, whereas with the estimators, \codename reduces this to 5 minutes - a 99\% reduction. Moreover, the profiling time of \codename with the estimators grows only marginally as the number of variants increases, demonstrating its scalability.

\textbf{Processor placement optimization:}
Figure~\ref{fig:placement_order_throughput} shows the inference throughput under different processor orders. Throughput varies widely, ranging from 110 to 211 on the desktop, 72 to 146 on the laptop, and 109 to 195 on the Jetson Orin - up to 2$\times$ difference. The optimal order also differs across testbeds: G–C–N on the desktop, G–N–C on the laptop, and G–C on the Jetson Orin. These results underscore the importance of optimized processor placement offered by \codename. With its \textit{Sparsity-Aware Optimizer}, \codename automatically selects the best order to maximize throughput.

\begin{figure*}[t!]
  \centering
  \begin{subfigure}[t]{0.32\linewidth}
    \centering
    \includegraphics[width=\linewidth]{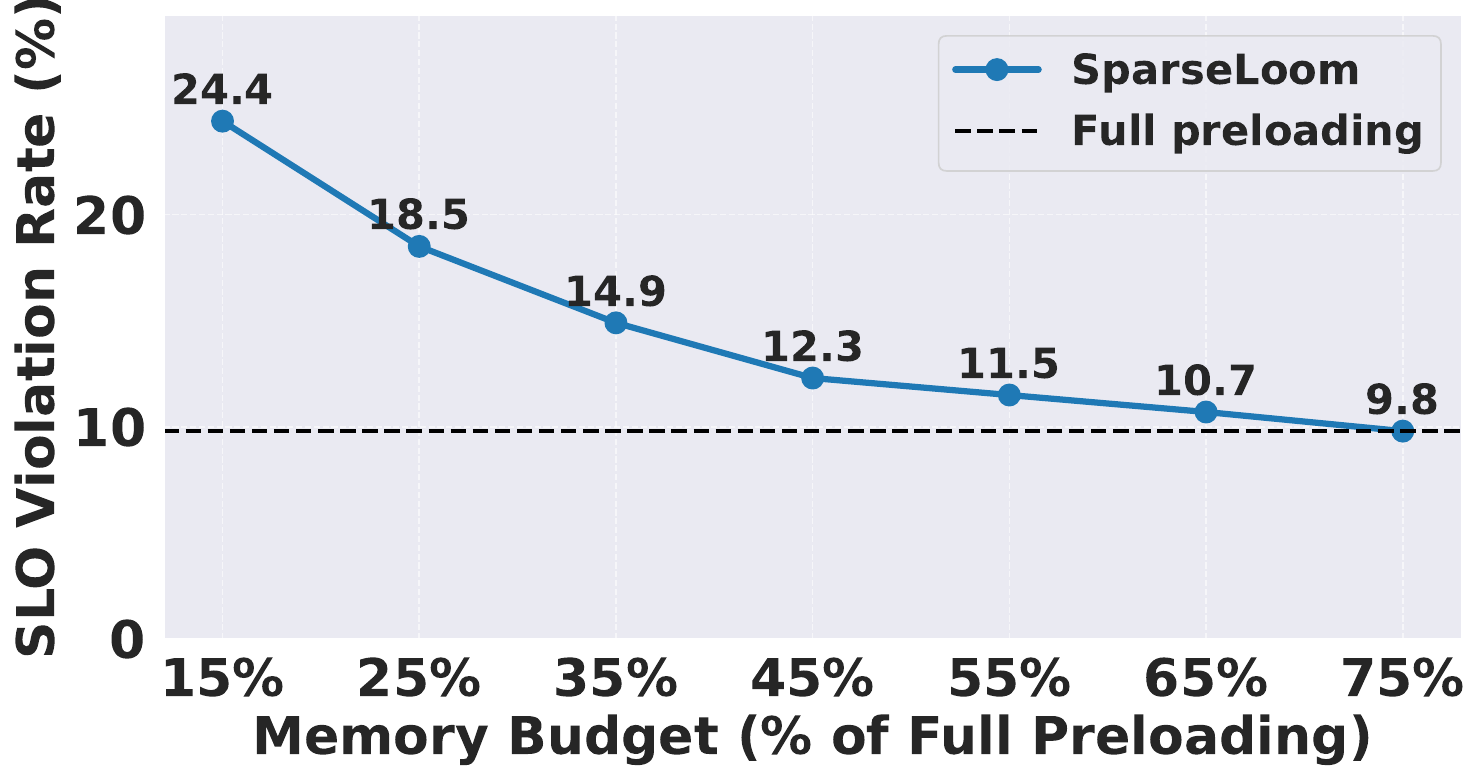}
    \caption{Desktop}
    \label{fig:Preloading_slo_desktop}
  \end{subfigure}
  \hfill
  \begin{subfigure}[t]{0.32\linewidth}
    \centering
    \includegraphics[width=\linewidth]{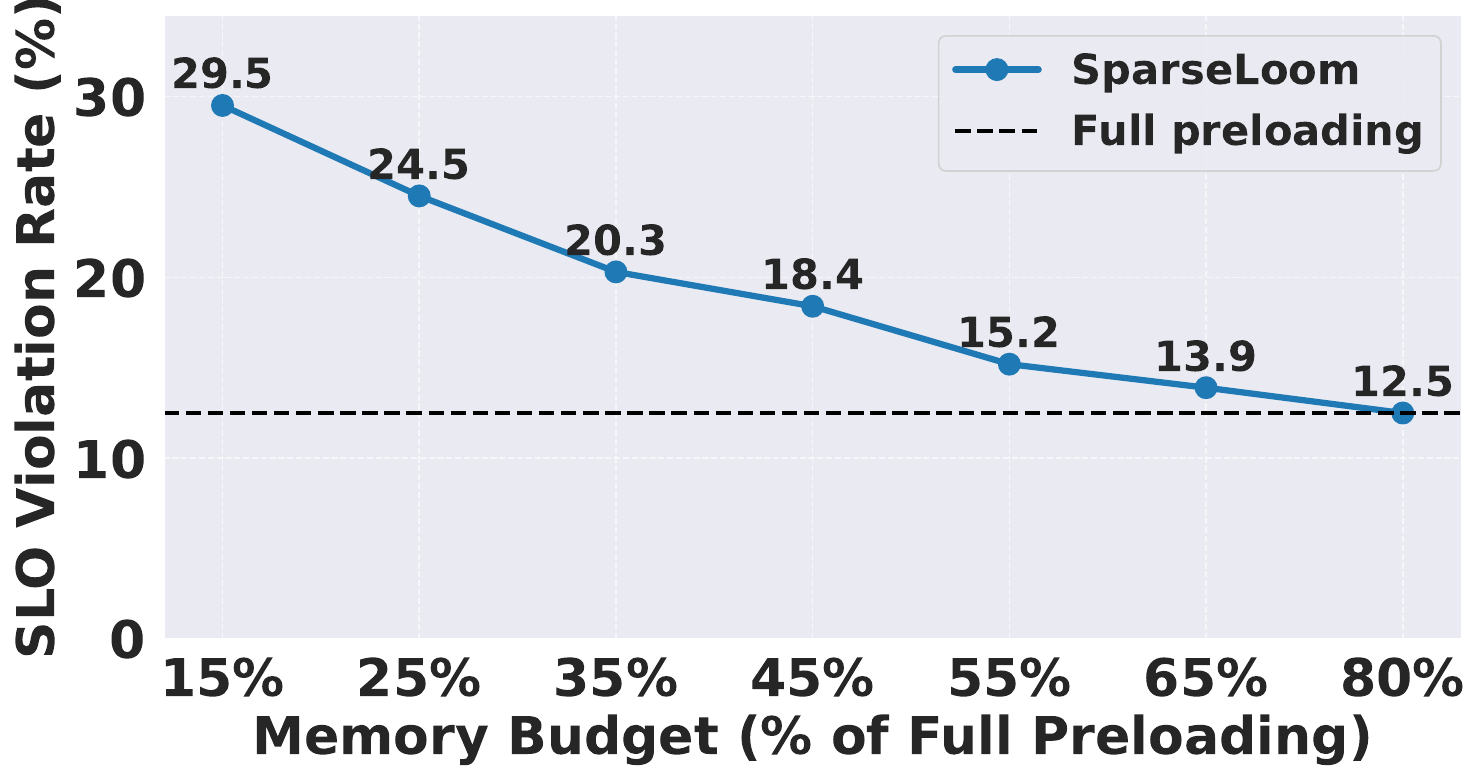}
    \caption{Laptop}
    \label{fig:Preloading_slo_laptop}
  \end{subfigure}
  \hfill
  \begin{subfigure}[t]{0.32\linewidth}
    \centering
    \includegraphics[width=\linewidth]{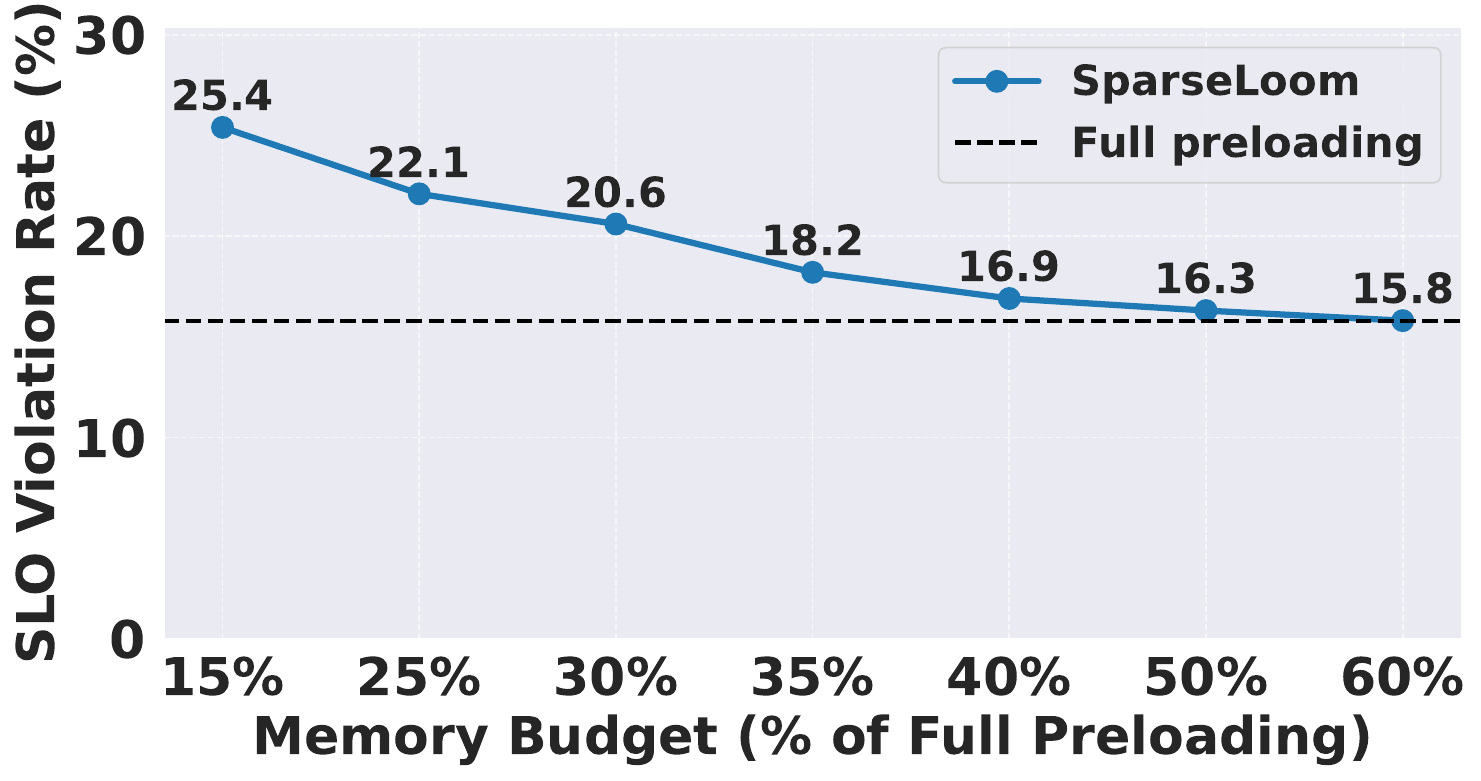}
    \caption{Jetson Orin}
    \label{fig:Preloading_slo_jetson}
  \end{subfigure}
  \caption{Impact of memory budget (percentage of memory required for fully preloading all subgraphs) on SLO violation. \codename cuts memory usage by average of 28\% without increasing the violation rate.}
  \label{fig:Preloading_slo_evaluation}
\end{figure*}

\textbf{Subgraph preloading:}
We further evaluate the \textit{Hot-Sub\\graphs Preloader} of \codename under different memory budgets. Figure~\ref{fig:Preloading_slo_evaluation} reports the SLO violation rates of \codename with varying memory budgets on three SoCs. We define the memory budget in this case as the proportion of memory required for fully preloading all subgraphs, referred to as full preloading. The lowest SLO violation rate on a given hardware platform will be achieved under full preloading. As expected, with smaller memory budgets the SLO violation rates increase in \codename, since fewer stitched variants can be preloaded. However, with only 15\% of the full preloading memory, \codename achieves violation rates of 24.4\%, 29.5\%, and 25.4\% on desktop, laptop, and Jetson Orin, already surpassing AV methods even with full memory. At a 55\% budget, it delivers comparable SLO violation rates within 2.7\% of full preloading. To reach the same violation rates as full preloading, \codename reduces memory usage by 25\%, 20\%, and 40\% on desktop, laptop, and Jetson Orin, respectively. These results demonstrate the effectiveness of the \textit{Hot-Subgraphs Preloader} in substantially reducing memory usage while maintaining comparable SLO performance.

\subsection{Discussion}
\textbf{Inter-processor execution overhead} in \codename includes (i) the communication cost of transferring intermediate activations from one processor to the next, and (ii) the format-conversion cost arising from differences in numerical precision. For example, when a stitched variant is executed in a GPU–CPU–NPU sequence, where the GPU uses FP32 precision and the CPU and NPU use INT8, the intermediate activations must be transferred from the GPU to the CPU and then from the CPU to the NPU, and the output precision must be converted from FP32 to INT8. Thus, both communication and format-conversion overheads are incurred.

We empirically measure that this inter-processor execution latency accounts for nearly 5\% of the total inference latency. The processors used in our experiments adopt a unified memory architecture~\cite{intel-265k,intel-135u,nvidia-orin}, and the format conversions are relatively lightweight and automatically handled by the inference engine.

\textbf{The number of partitioned subgraphs} also impacts inference latency. Increasing the number of subgraphs enables more fine-grained scheduling. However, the degree of parallelism does not scale indefinitely, as it is bounded by the number of available processors. In our experiments, the number of model subgraphs is equal to the number of processors. This configuration is also commonly adopted in existing multi-DNN inference systems~\cite{hetero2pipe2025,karatzas2023omniboost,karatzas2025rankmap}.

\textbf{Impact of dynamic voltage and frequency scaling (DVFS).}
In all our experiments, we used the default DVFS policy with no  manual modifications. However, we do not observe DVFS actively adjusting processor frequencies during execution. When DVFS is triggered, it may reduce processor frequencies and consequently increase the inference latency of all variants. This affects all baselines, not only \codename. To mitigate this effect, \codename can incorporate a simple mechanism to re-profile all variants for identifying the optimal processor order under the new DVFS conditions.

\textbf{Applicability of \codename to large foundation models.}
\codename focuses on edge-sized DNNs, including CNNs and transformer-based models. Even using lightweight DNNs, multi-DNN parallel execution already leads to high SLO violation rates due to the computational limits of edge resources.
In contrast, large foundation models (e.g., large language models) remain challenging to be deployed on edge SoCs and are  served as a single multimodal model rather than multiple DNNs. Optimizing the execution efficiency of such models is outside the scope of \codename; they have fundamentally different bottlenecks, dominated by extremely large memory requirements and memory-access costs~\cite{xue2024powerinfer}.
\section{Related Work}
\label{sec:related work}
We compare \codename with three classes of multi-DNN inference systems and variant-generation methods, with Appendix~\ref{appendix:features} summarizing their key features.


\textit{\textbf{Class~1}-Task-level scheduling systems on a single processor:}  
These systems optimize the execution order of multiple DNN tasks on a single processor to meet latency constraints.  
Representative examples include Pipe-it~\cite{wang2019high}, Pantheon~\cite{han2024pantheon}, and REEF~\cite{han2022microsecond}.  
They typically adopt techniques, such as priority-based scheduling and deadline-aware ordering, to maximize utilization on limited compute resources.  
A key limitation is their lack of support for heterogeneous processors, which results in compute resource underutilization.  

\textit{\textbf{Class~2}-Subgraph placing systems on heterogeneous processors:}  
These systems partition DNNs into subgraphs and assign to heterogeneous processors.  
Representative systems include Hetero$^2$Pipe~\cite{hetero2pipe2025}, Band~\cite{jeong2022band}, and OmniBoost~\cite{karatzas2023omniboost}.  
By executing subgraphs on different processors in parallel, these methods improve the system throughput.  
However, they assume a single model per task and therefore lack support for multiple variant selection to adapt to diverse SLOs.  

\textit{\textbf{Class~3}-Multiple variant selecting systems for diverse SLOs:}  
These systems enable inference frameworks to select from multiple model variants to meet diverse accuracy-latency requirements.  
Representative works include Tango~\cite{taufique2024tango}, \allowbreak ESIM~\cite{sen2024elastically}, and NestDNN~\cite{fang2018nestdnn}.  
Typically, they prepare a set of model variants with different accuracy and latency performance, and the selection is guided by profiling data, historical performance, and heuristic rules to balance latency and accuracy.  
However, their ability to support diverse SLOs is often limited by the few available model variants.

\textbf{Variant-generation methods} such as adaptive sparsity~\cite{sanh2020movement} and per-layer pruning~\cite{ro2021autolr} can generate a small number of model variants (typically tens) similar to the model stitching used by \codename. However, these approaches require additional computational resources and calibration datasets for retraining. In contrast, \codename is data- and training-free, and can significantly enlarge the variant space. Additionlly, variants generated by these methods may satisfy SLOs when evaluated individually but violate them under multi-DNN execution. \codename goes beyond variant-generation approaches by providing a holistic system for variant generation, selection, and placement.

\section{Conclusion}
\label{sec:conclusion}
In this paper, we proposed \codename, the first multi-DNN inference system that leverages model stitching to expand the model variant space and reduce SLO violation rates. To efficiently support stitching, \codename develops a lightweight performance profiler, a throughput-optimized processor placement module, and a memory-efficient subgraph preloader. Evaluations across diverse edge SoCs show that \codename reduces SLO violations by up to 74\%, improves throughput by up to 2.31$\times$, and reduces memory overhead by an average of 28\% compared to state-of-the-art systems.



\bibliographystyle{ACM-Reference-Format}
\bibliography{main}

\newpage
\thispagestyle{empty}
\appendix

\section{Sparse Variants}
\label{appendix:variants} 
\textbf{Variants used:} Table~\ref{tab:sparsity_variants_one} presents the complete set of sparse variants used in our evaluation. For each task, we construct a sparse model zoo comprising ten variants to capture diverse sparsity patterns. The set includes one dense base model, one quantized model (INT8) for Intel SoCs, two quantized models (FP16 and INT8) for Jetson Orin~\cite{wu2020easyquant}, and the remaining variants are obtained by pruning~\cite{eccles2024dnnshifter} - eight pruned models for Intel SoCs and seven for Jetson Orin - the dense base model.

\textbf{Creating the variants:}
On Intel SoCs, we use OpenVINO’s Neural Network Compression Framework (NNCF) to generate sparse variants using structured pruning (that changes the architecture), unstructured pruning (using zero-masking)~\cite{cheng2024survey}, and quantization. On NVIDIA SoCs, we employ the ONNX framework to produce pruned (structured) and quantized variants. These compression frameworks are widely adopted in the real-world to generate sparse variants suited for edge deployments in the order of a few minutes per model. All compression methods are applied after training, and therefore, require hundreds of calibration samples. These variants are stored in persistent storage (e.g., on disk), totaling 7.6 GB for all sparse model zoos across all four tasks in our experiments.

\textbf{Software/Hardware support for different compression types:}
Structured pruning is hardware- and software-agnostic. Unstructured pruning, however, relies on sparse-acceleration software (e.g., DeepSparse~\cite{DeepSparse}) and is independent of the underlying hardware platform. Quantization requires both software and hardware support.
Most representative SoCs (e.g., the NVIDIA Jetson series) support 8-bit quantization. In our experiments, the Intel Ultra 5 and Ultra 7 platforms support all three compression types, whereas the Jetson Orin does not provide support for unstructured pruning.

\begin{table}[htbp]
\centering
\caption{Sparse variants on each platform. \emph{Sparse Pattern} indicates sparsity level and numerical precision of the model parameters. Note: $x$\% sparsity means $x$\% weights are pruned.}
\label{tab:sparsity_variants_one}
\begin{tabular}{P{2.8cm} P{2.8cm} P{2cm}}
\toprule
\textbf{Platform} & \textbf{Variant Type} & \textbf{Sparse Pattern} \\
\midrule
\multirow{9}{*}{\makecell{Intel SoCs \\ (Desktop / Laptop)}} 
    & Base model        & Dense (FP32) \\
    & Quantized model   & Dense (INT8) \\
    \cmidrule(lr){2-3}
    & \multirow{6}{*}{Unstructured pruning} 
        & 90\%, FP32 \\
    &                   & 85\%, FP32 \\
    &                   & 80\%, FP32 \\
    &                   & 75\%, FP32 \\
    &                   & 70\%, FP32 \\
    &                   & 65\%, FP32 \\
    \cmidrule(lr){2-3}
    & \multirow{2}{*}{Structured pruning} 
        & 40\%, FP32 \\
    &                   & 50\%, FP32 \\
\midrule
\multirow{10}{*}{NVIDIA Jetson} 
    & Base model        & Dense (FP32) \\
    & Quantized model   & Dense (FP16) \\
    &                   & Dense (INT8) \\
    \cmidrule(lr){2-3}
    & \multirow{7}{*}{Structured pruning} 
        & 20\%, FP32 \\
    &                   & 30\%, FP32 \\
    &                   & 35\%, FP32 \\
    &                   & 40\%, FP32 \\
    &                   & 45\%, FP32 \\
    &                   & 50\%, FP32 \\
    &                   & 55\%, FP32 \\
\bottomrule
\end{tabular}
\end{table}

\clearpage
\section{Notation}
\label{appendix:notation} 

The notation used in the paper is summarized in Table~\ref{tab:notation}.

\begin{table}[htbp]
\centering
\caption{Summary of notation.}
\label{tab:notation}
\begin{tabular}{ll}
\Xhline{2\arrayrulewidth}
\textbf{Notation} & \textbf{Description} \\
\Xhline{2\arrayrulewidth}
$t, T$ & Task index and total number of tasks \\
$i, V$ & Variant index and variants per task \\
$j, S$ & Subgraph index and subgraphs per variant \\
$k$ & Stitched variant index (total $V^S$ variants) \\
\hline
$v^{t,i}$ & Variant $i$ in task $t$ \\
$s_j^{t,i}$ & Subgraph $j$ of variant $i$ in task $t$ \\
$\tilde{v}^{t,k}$ & Stitched variant $k$ in task $t$\\
$\tilde{s}_j^{t,k}$ & Subgraph $j$ of stitched variant $k$ in task $t$ \\
$\tilde{v}^{t,*}$ & Optimal stitched variants of task $t$\\
\hline
$p_j$ & Processor assigned to subgraph $j$ \\
$\vec{p}$ & Placement order $\{p_1, \dots, p_S\}$ \\
$\vec{p}^*$ & Optimal placement order for all tasks \\
$\mathit{\Omega}$ & Set of all $P!$ placement orders \\
$\mathit{Lat}(s_j^{t,i}, p_j)$ & Latency of $s_j^{t,i}$ on $p_j$ \\
$\mathit{Lat}(\tilde{v}^{t,k}, \vec{p})$ & Latency of $\tilde{v}^{t,k}$ for $\vec{p}$ \\
$A(\tilde{v}^{t,k})$ & Accuracy of $\tilde{v}^{t,k}$ \\
\hline
$SLO^{t}_{acc}$ & Accuracy SLO of task $t$ \\
$SLO^t_{lat}$ & Latency SLO of task $t$ \\
$\Theta^{t}$ & Variants satisfying both $SLO^{t}_{acc}$ and $SLO^{t}_{lat}$ \\
\hline
$Mem(s_j^{t,i})$ & Memory cost of $s_j^{t,i}$ \\
$Mem_{budget}$ & Total memory budget \\
$\Phi^{t}$ & Cached subgraphs for task $t$ \\
$\sigma$ & An SLO configuration\\
$\Psi$ & Set of SLO configurations \\
\Xhline{2\arrayrulewidth}
\end{tabular}
\end{table}

\section{Comparison of existing methods with \codename}
\label{appendix:features} 

Table~\ref{table:works} compares existing classes of multi-DNN inference systems, variant-generation methods, and \codename across key capabilities. Class~1 systems optimize task scheduling on a single processor, Class~2 systems focus on subgraph placement on heterogeneous processors, and Class~3 systems support model variant selection for diverse SLOs. However, these three classes provide only a single model or tens of variants and lack support for sparsity-aware placement and memory-efficient preloading. Variant-generation methods produce fine-grained sparse variants but require calibration data or retraining and remain limited in scale. In contrast, \codename is data- and training-free, supports thousands of stitched variants, and integrates variant generation, selection, sparsity-aware placement, and memory-efficient preloading to enable efficient multi-DNN inference.

\begin{table*}[tp]
\centering
\caption{Key features of different classes of multi-DNN inference systems, variants generation
methods and \codename. 
\textit{Class~1}: systems that optimize task-level scheduling on a single processor, 
\textit{Class~2}: systems that optimize subgraph placement on heterogeneous processors, and 
\textit{Class~3}: systems that support the selection of multiple variants for diverse SLOs. Variant-generation methods generates fine-grained sparse variants.}
\begin{tabular}{P{6cm} P{1.5cm} P{1.5cm} P{1.5cm} P{3.5cm} P{1.8cm}}
\Xhline{2\arrayrulewidth}
\textbf{Features}
& \textbf{Class~1} \space  ~\cite{han2024pantheon,wang2019high,han2022microsecond}
& \textbf{Class~2} \space~\cite{hetero2pipe2025,jeong2022band,karatzas2023omniboost}
& \textbf{Class~3} \space ~\cite{taufique2024tango,sen2024elastically,fang2018nestdnn}
 & \textbf{Variants Generation Methods}~\cite{sanh2020movement,ro2021autolr}
 & \textbf{\codename} \\
\Xhline{2\arrayrulewidth}

Data- and training-free variant generation
& \xmark & \xmark & \xmark 
& \xmark 
& \cmark \\
\hline

Multiple variants selection 
& \xmark & \xmark & \cmark 
& \xmark 
& \cmark \\

\hline

\# model variants supported 
& 1 & 1 & Scale of 10 
& Scale of 10 
& Scale of 1000 \\

\hline

Heterogeneous subgraph placement 
& \xmark & \cmark & \xmark 
& \xmark 
& \cmark \\

\hline

Optimized sparsity-aware placement 
& \xmark & \xmark & \xmark 
& \xmark 
& \cmark \\

\hline

Memory-efficient preloading 
& \xmark & \xmark & \xmark 
& \xmark 
& \cmark \\

\Xhline{2\arrayrulewidth}
\end{tabular}
\label{table:works}
\end{table*}

\section{Comparing \codename and Baselines Under Accuracy- and Latency-Guaranteed SLOs}
\label{appendix:experiments_slo} 

We define \textit{accuracy-guaranteed SLOs} as those where the accuracy requirement is fixed to the highest value across all variants while varying the latency threshold. Conversely, \textit{latency-guaranteed SLOs} fix the latency requirement to the lowest value while varying the accuracy threshold.

For example, suppose the accuracy range across all variants is $[85\%, 92\%]$ and the latency range is $[50\,\mathrm{ms}, 120\,\mathrm{ms}]$. For accuracy-guaranteed SLOs, we fix the accuracy requirement at $92\%$ and uniformly generate five latency thresholds between $50$ and $120\,\mathrm{ms}$ (e.g., $\{50, 67.5, 85, 102.5, 120\}\,\mathrm{ms}$). Conversely, for latency-guaranteed SLOs, we fix the latency requirement at $50\,\mathrm{ms}$ and uniformly generate five accuracy thresholds between $85\%$ and $92\%$ (e.g., $\{85\%, 86.75\%, 88.5\%$, $90.25\%, 92\%\}$).

These SLOs capture scenarios that require \textit{no accuracy compromise} or \textit{no latency compromise}, reflecting applications that prioritize either accuracy or latency over the other.

\begin{figure*}[ht]
  \centering
  \begin{subfigure}[t]{0.32\linewidth}
    \centering
    \includegraphics[width=\linewidth]{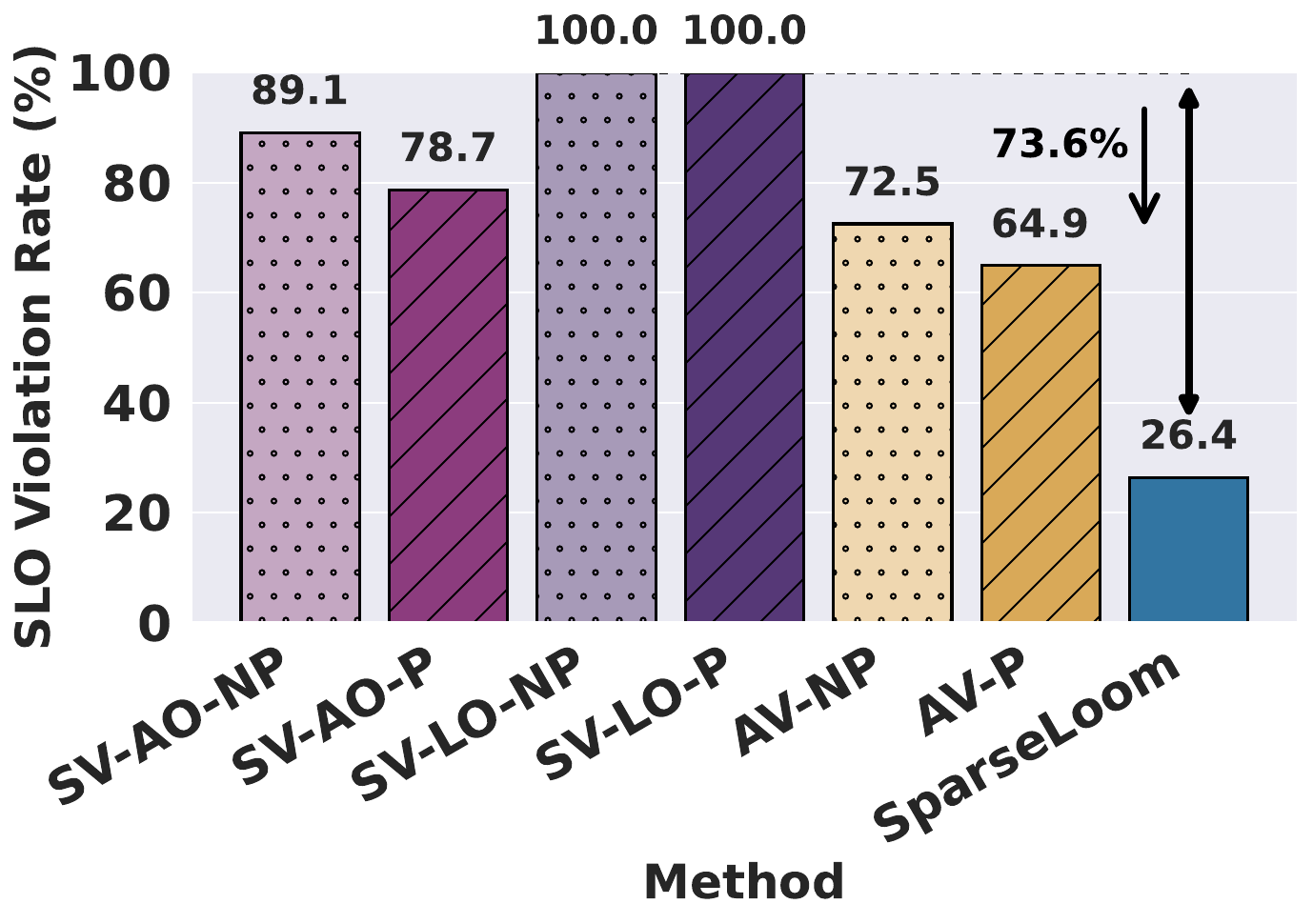}
    \caption{Desktop}
  \end{subfigure}
  \hfill
  \begin{subfigure}[t]{0.32\linewidth}
    \centering
    \includegraphics[width=\linewidth]{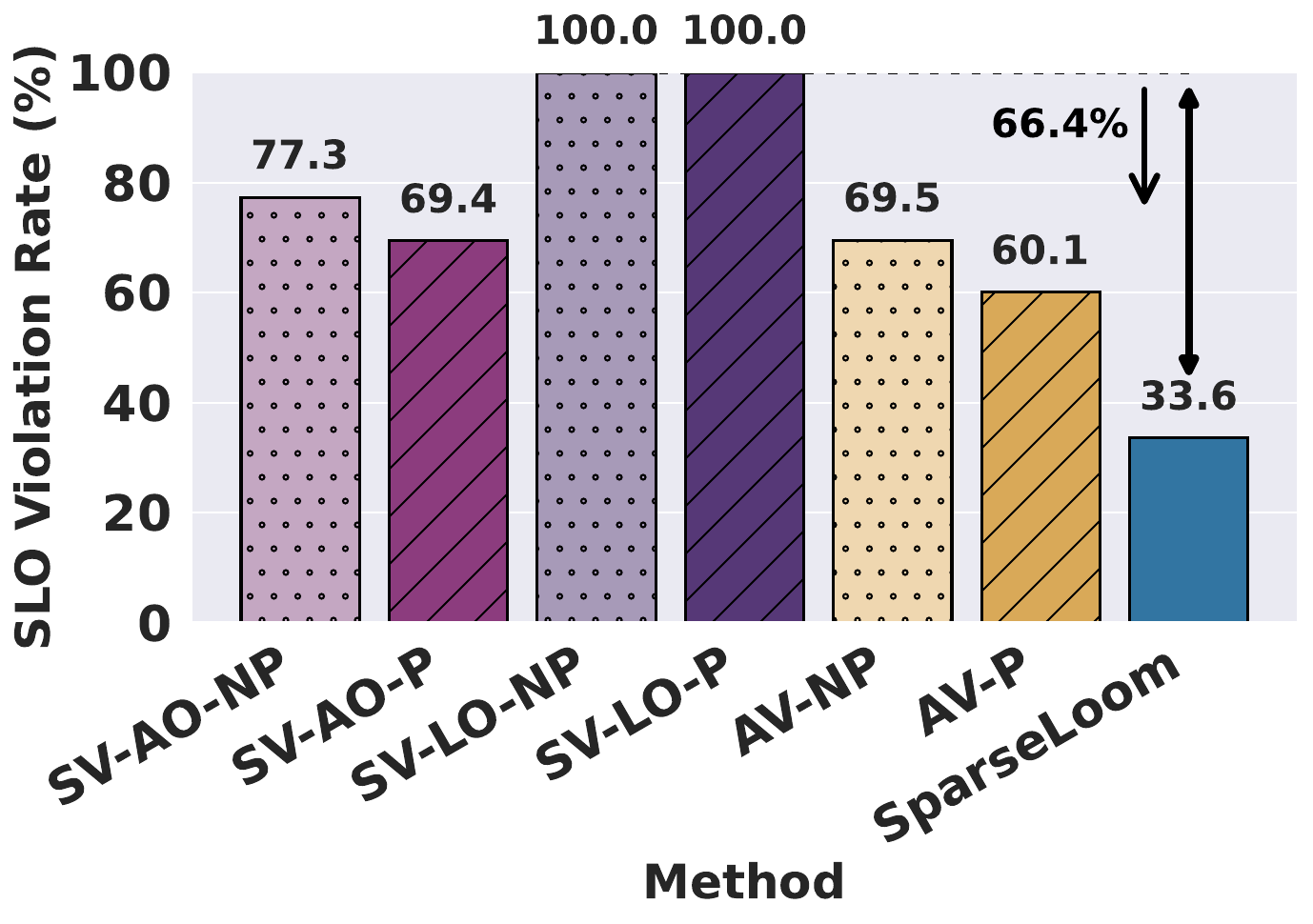}
    \caption{Laptop}
  \end{subfigure}
  \hfill
  \begin{subfigure}[t]{0.32\linewidth}
    \centering
    \includegraphics[width=\linewidth]{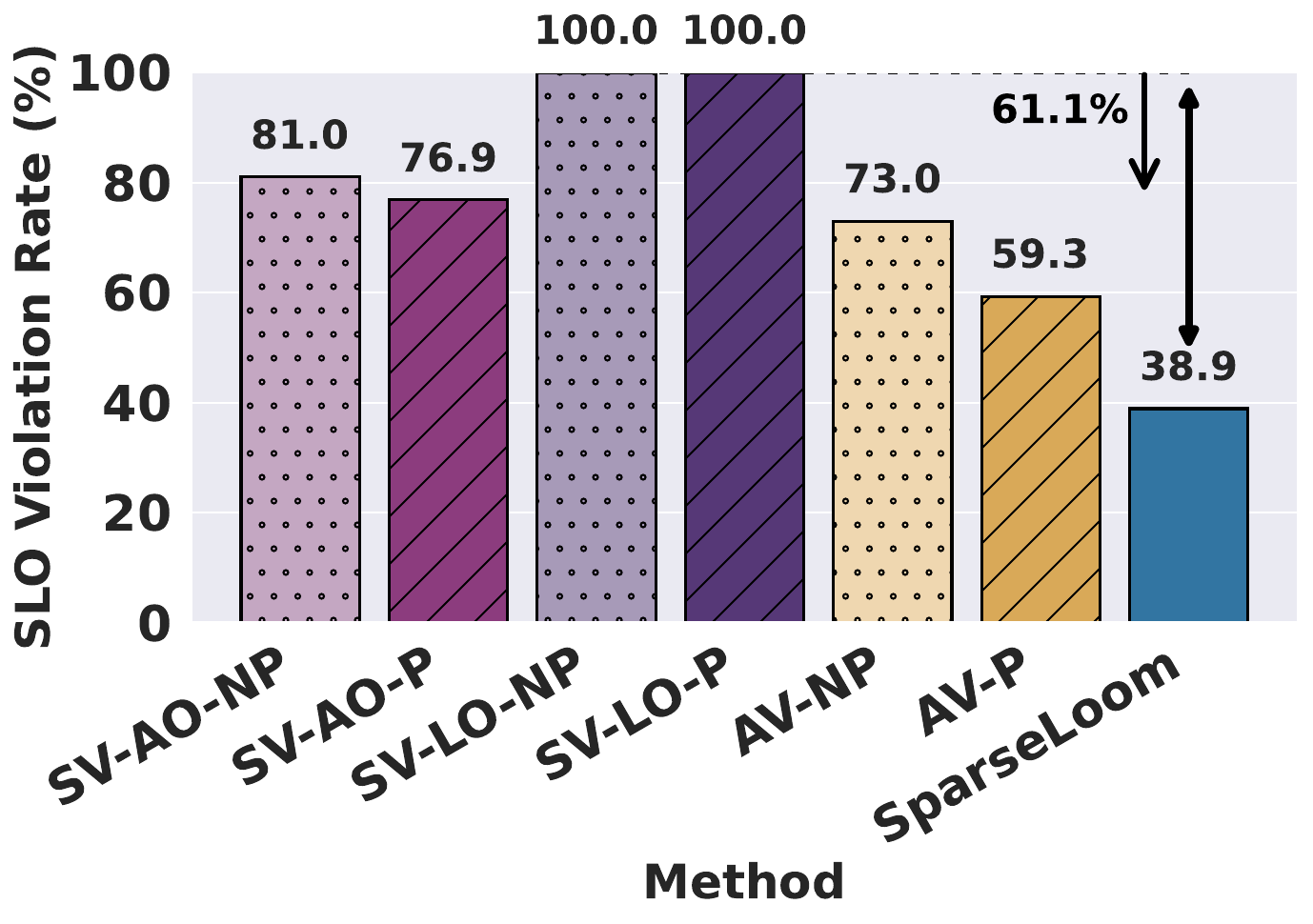}
    \caption{Jetson Orin}
  \end{subfigure}
  \caption{SLO violation rates of \codename and baselines across three SoCs under accuracy-guaranteed SLOs. \codename consistently achieves lower violation rates than all baselines, with reductions of up to 73.6\%.}
  \label{fig:slo_accuracy}
\end{figure*}

\begin{figure*}[ht]
  \centering
  \begin{subfigure}[t]{0.32\linewidth}
    \centering
    \includegraphics[width=\linewidth]{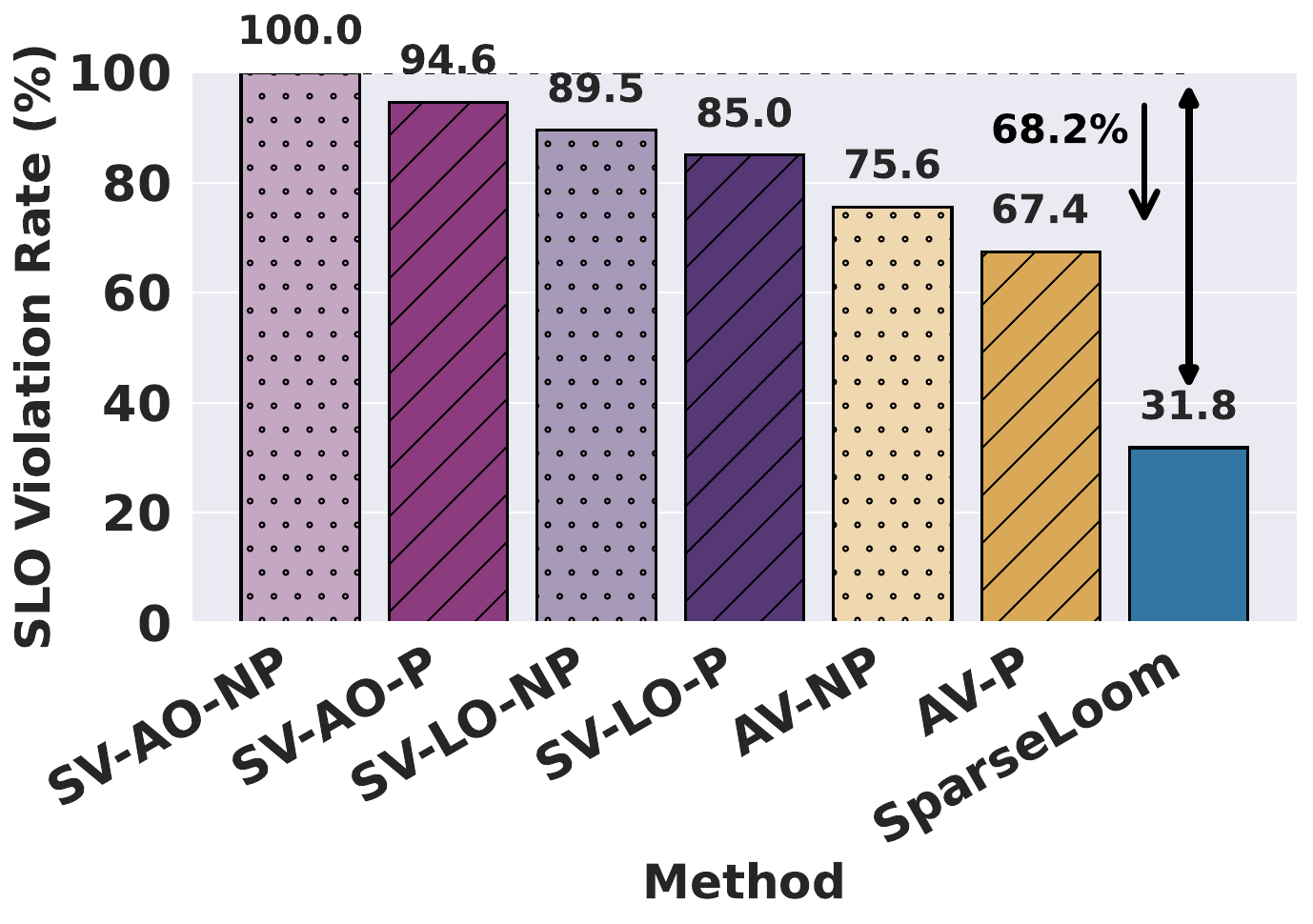}
    \caption{Desktop}
  \end{subfigure}
  \hfill
  \begin{subfigure}[t]{0.32\linewidth}
    \centering
    \includegraphics[width=\linewidth]{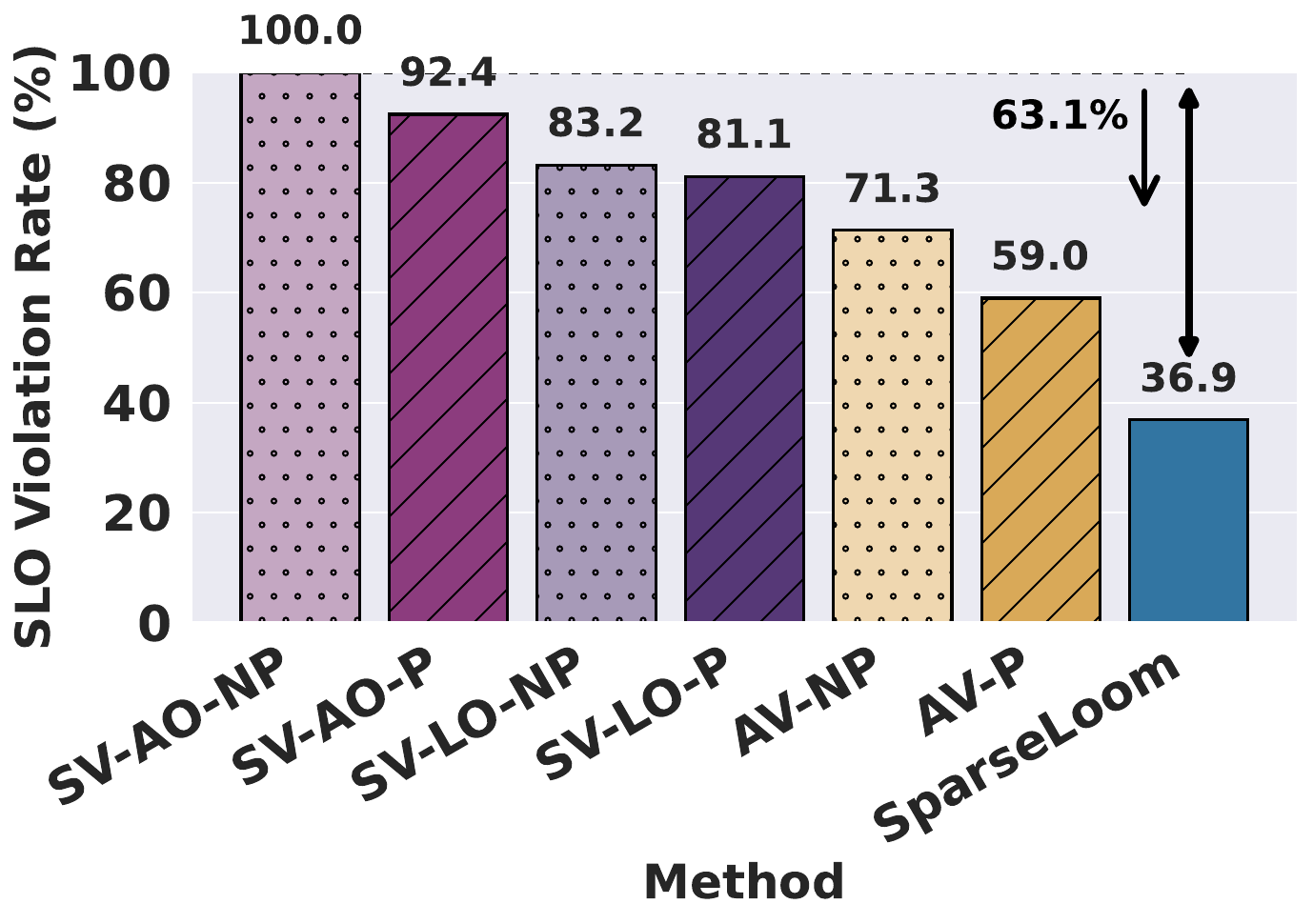}
    \caption{Laptop}
  \end{subfigure}
  \hfill
  \begin{subfigure}[t]{0.32\linewidth}
    \centering
    \includegraphics[width=\linewidth]{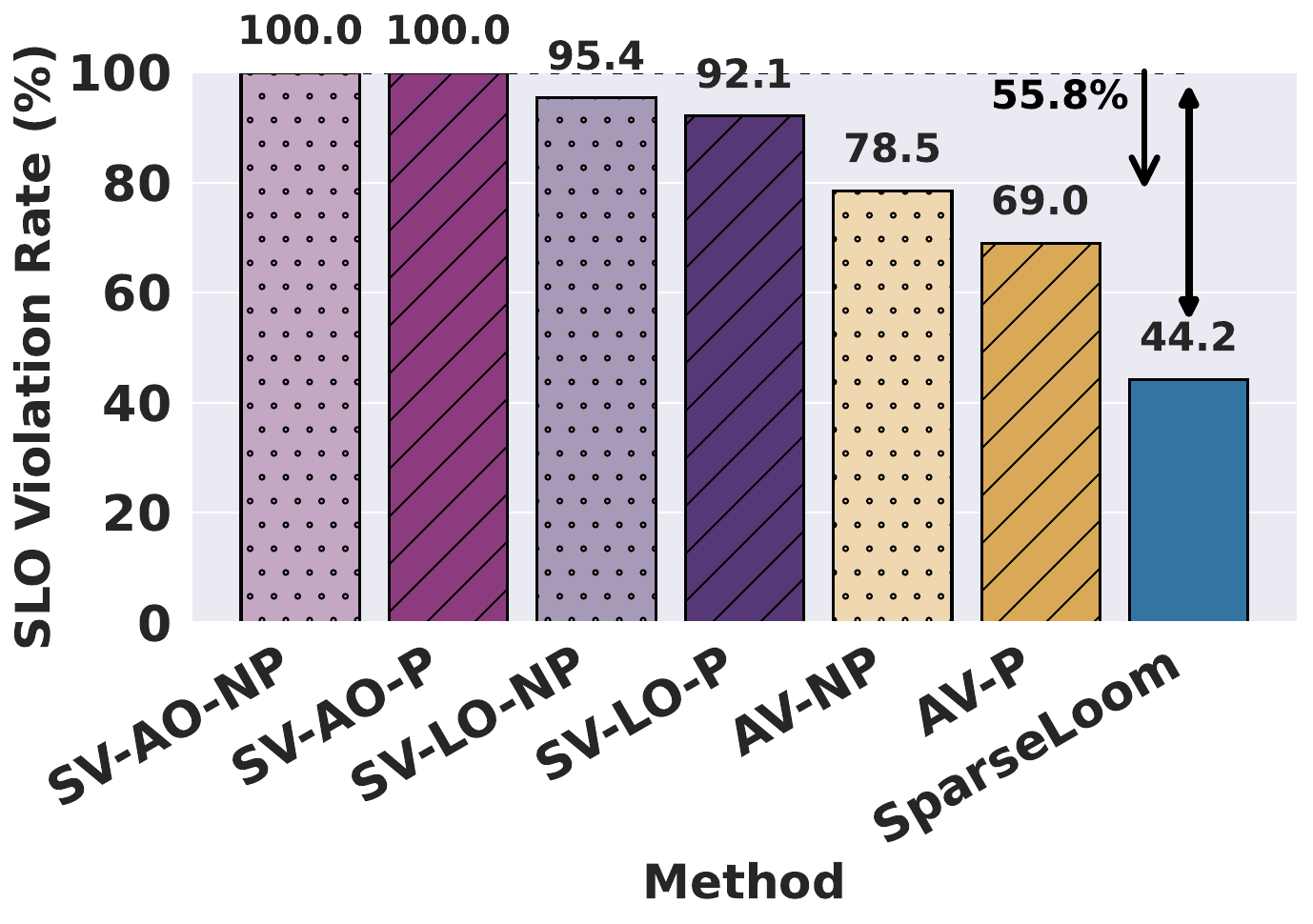}
    \caption{Jetson Orin}
  \end{subfigure}
  \caption{SLO violation rates of \codename and baselines across three SoCs under latency-guaranteed SLOs. \codename consistently achieves lower violation rates than all baselines, with reductions of up to 68.2\%..}
  \label{fig:slo_latency}
\end{figure*}

Figure~\ref{fig:slo_accuracy} and Figure~\ref{fig:slo_latency} show the SLO violation rates of \codename and all baselines on the three hardware platforms presented in the Evaluation section. Under both accuracy-guaranteed and latency-guaranteed SLOs, \codename consistently reduces SLO violation rates by up to 73.6\% and 68.2\%, respectively. This demonstrates that \codename not only lowers SLO violation rates on average across all SLOs, but also remains effective in scenarios with stringent accuracy or latency requirements.

\end{document}